\documentclass{aa}
\usepackage{amsmath}
\usepackage{graphicx}
\usepackage[utf8]{inputenc}
\usepackage{natbib}

\usepackage{txfonts}
\graphicspath{{Graficos/}}
\usepackage[normalem]{ulem}
\usepackage{color}

\begin{document}

   \title{New multi-part collisional model of the Main Belt: \\ The contribution to near-earth asteroids}    
   \author{P. S. Zain\inst{1,2}\thanks{pzain@fcaglp.unlp.edu.ar}, 
           G. C. de El\'{\i}a\inst{1,2}, 
           \and R. P. Di Sisto\inst{1,2}
           }

   \offprints{Patricio Zain
    }
  \institute{Instituto de Astrof\'{\i}sica de La Plata, CCT La Plata-CONICET-UNLP \\                                                                                       
   Paseo del Bosque S/N (1900), La Plata, Argentina                                                                                                                         
   \and Facultad de Ciencias Astron\'omicas y Geof\'{\i}sicas, Universidad Nacional de La Plata \\                                                                           
   Paseo del Bosque S/N (1900), La Plata, Argentina 
                }

   \date{Received / Accepted}


\abstract
{}
{We developed a six-part collisional evolution model of the main asteroid belt (MB) and used it to study the contribution of the different regions of the MB to the near-earth asteroids (NEAs).}
{We built a statistical code called \texttt{ACDC} that simulates the collisional evolution of the MB split into six regions (namely Inner, Middle, Pristine, Outer, Cybele and High-Inclination belts) according to the positions of the major resonances present there ($\nu_{6}$, 3:1J, 5:2J, 7:3J and 2:1J). We consider the Yarkovsky effect and the mentioned resonances as the main mechanism that removes asteroids from the different regions of the MB and delivers them to the NEA region. We  calculated the evolution of the NEAs coming from the different source regions by considering the bodies delivered by the resonances and mean dynamical timescales in the NEA population. }
{Our model is in agreement with the major observational constraints associated with the MB, such as the size distributions of the different regions of the MB and the number of large asteroid families. It is also able to reproduce the observed NEAs with $H<16$ and agrees with recent estimations for $H<20$, but deviates for smaller sizes. We find that most sources make a significant contribution to the NEAs; however the Inner and Middle belts stand out as the most important source of NEAs followed by the Outer belt. The contributions of the Pristine and Cybele regions are minor. The High-Inclination belt is the source of only a fraction of the actual observed NEAs with high inclination, as there are dynamical processes in that region that enable asteroids to increase and decrease their inclinations.}
{}

\keywords{minor planets, asteroids: general -- methods: numerical -- methods: statistical}
\authorrunning{P. S. Zain et al.}
\titlerunning{New multi-part collisional model of the Main Belt: The contribution to near-earth asteroids}

\maketitle
\section{Introduction}

Asteroids are minor bodies located mostly in the inner Solar System and are considered to be the shattered remains of planetesimals, the bricks from which the planets were formed. There are millions of asteroids,  all with irregular shapes and sizes ranging from dust particles to diameters of hundreds of kilometers. The large majority of asteroids reside in the main asteroid belt (MB), a vast region in our Solar System located between the orbits of Mars and Jupiter, from $\sim 2$ AU to $\sim 3.4 $ AU from the Sun. Another population of asteroids worth studying are the near-Earth asteroids (NEAs), because they are able to cross the orbit of the Earth and therefore may represent a hazard for human civilization in case of an impact with our home planet. The NEAs are part of a bigger population called near-Earth objects (NEOs), which also includes comets. 

Since the pioneering paper by \cite{dohnanyi1969}, many studies have been published on the collisional evolution of the MB. In particular, \cite{PetitFarinella1993} and afterwards \cite{OBrien2005} considered empiric fragmentation laws derived from laboratory experiments performed by \cite{Nakamura1991}. Also, \cite{Morbidelli2009} constructed a statistical code named \texttt{BOULDER} using the results of smooth-particle hydrodynamics (SPH) simulations performed by \cite{BenzAsphaug} and \cite{Durda2007}. The works of \cite{Bottke2005a,Bottke2005b} performed a collisional evolution model to constrain the primordial size distribution of the MB, and the link between the collisional history and its dynamical excitation and depletion. All of the mentioned papers considered the whole MB as a single population. However, different regions of the belt may have different physical and dynamical properties. Therefore, a multiple-population model of the MB can lead to a more realistic representation of the collisional and dynamical processes that affect the MB. In particular, \cite{deElia2007} built a collisional and dynamical evolution model of the MB by splitting it into three regions: inner, middle, and outer belt, and studied the contribution of the different regions to the NEA population. More recently, \cite{Cibulkova2014} developed a six-population collisional evolution model of the MB and tested different fragmentation laws considering monolithic and rubble-pile bodies. The work of \cite{Cibulkova2014} also made a global exploration of the Yarkovsky effect, testing different thermal and physical models, and thus serving as starting points for future improvements of collisional models. 

The MB and NEA populations are closely connected by evolutionary processes and dynamical transport mechanisms. Intense collisional activity in the MB continuously shatters asteroids and injects a large quantity of material into the resonant regions via radiation forces like the Yarkovsky effect  \citep{Farinella1998b,Morbidelli2003}. Nowadays, it is widely accepted that the main escape routes from the MB and source of NEAs are the main resonances with the outer planets.  \cite{Gladman} showed that objects falling into resonances become NEAs in a few million of years. Recently, \cite{Granvik2016} performed $N$-body simulations aimed at locating the escape routes from the MB into the NEA region.  

Many works have been made to estimate the population of NEAs. \cite{Bottke2002} constructed a debiased orbital and absolute magnitude-frequency distribution (HFD) of NEOs by using $N$-body simulations, which was updated recently by \cite{Granvik2018}. Other papers estimated the HFD using collisional evolution models, as in the work of \cite{OBrien2005} and \cite{deElia2007}. Also, \cite{Harris2015}, \cite{SchunovaLilly2017}, and \cite{Tricarico2017} derived HFDs of the NEAs using observational methods. Nowadays, the observed HFD is thought to be complete for $H<16$ \citep{Tricarico2017}.

Here, we present a new multi-population code called \texttt{ACDC} (Asteroid Collisions and Dynamic Computation) based on the prescriptions of the \texttt{BOULDER} code \citep{Morbidelli2009} and the work of \cite{Cibulkova2014}. \texttt{ACDC} is a statistical code that simulates the collisional evolution of the MB split into six regions bounded by the major resonances present. We included the action of the Yarkovsky effect and resonances as the mechanism that removes asteroids from the MB. The inclusion the Yarkovsky effect in a multi-population model is a big improvement with respect to previous works, because the Yarkovsky effect greatly influences the collisional evolution of the MB, and enables us to perform a new study of the link between the MB and the NEAs. 

The main goal of this paper is to develop a six-part collisional evolution model of the MB and use it to obtain a HFD of the NEA region, and also determine the source regions of the NEAs. This study is an attempt to answer the following questions: Where do the NEAs come from? How many NEAs come from the different regions of the MB? Is there a particular region in the MB that provides more NEAs than the others? Is there a specific region that delivers bigger asteroids? 

The paper is structured as follows. In Section 2, the partition of the MB in six regions is presented. In Section 3, we fully describe the collisional evolution model of the MB and the Yarkovsky effect. In Section 4, we describe our model for the NEA dynamical evolution. We present our results for the initial and final size--frequency distributions (SFDs) of the six regions of the MB, the HFD of NEAs, and the contribution of the different regions in Section 5. Finally, we present our conclusions  and discuss our findings in Sect. 6. 

\section{Partition of the asteroid belt}

In the past, most studies on collisional evolution, such as \citet{Bottke2005a} and \citet{OBrien2005}, were carried out assuming the MB as a whole, considering one single probability of collision and impact speed for all asteroids. However, different regions of the belt may have different physical and dynamical properties such as number of objects, composition, distance to the Sun, presence of resonances, and so on. 

\begin{figure}[htp]
\centering
\includegraphics[width=8cm]{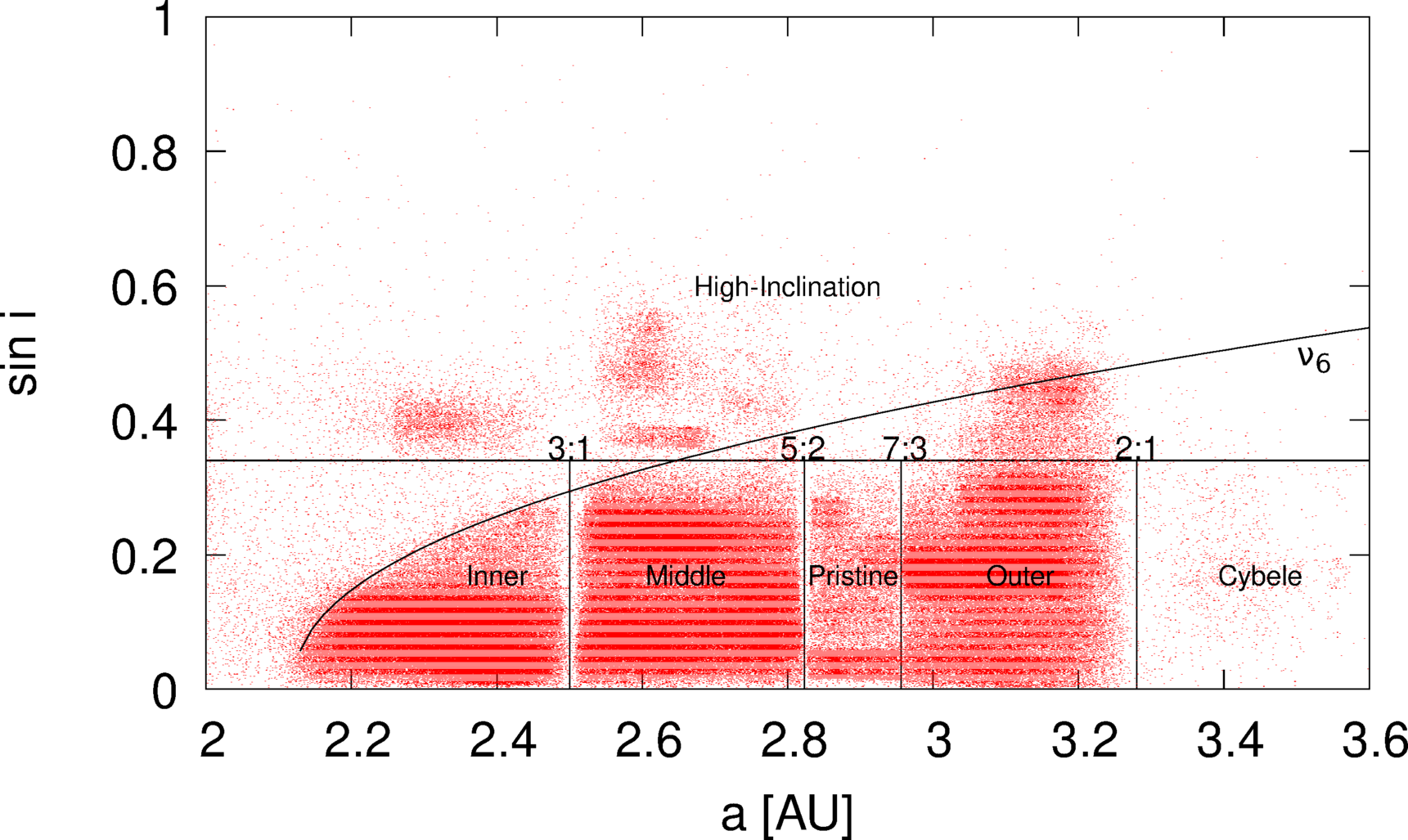}
\caption{Main asteroid belt plotted in semimajor axis $a$ vs. inclination $I$. The six defined regions are Inner, Middle, Pristine, Outer, Cybele, and High-Inclination, which are separated by the positions of the major resonances in the asteroid belt. The resonances considered are $\nu_{6}$, 3:1, 5:2, 7:3 and 2:1. The curve denoting the bond of the $\nu_{6}$ resonance was plotted according to \cite{Morbidelli2003}. Data were obtained from the Minor Planet Center.}
    \label{fig:MB}
\end{figure}

In order to make a more realistic model of the interactions between the different parts of the MB, we follow \cite{Cibulkova2014}  and split the MB into six regions, or populations, separated by the major mean-motion resonances (MMR) with Jupiter and the $\nu_{6}$ secular resonance with Saturn. The asteroid belt and the locations of the MMR are plotted in semimajor-axis versus inclination in Fig.~\ref{fig:MB} using the online data from the Minor Planet  Center \footnote{\texttt{https://www.minorplanetcenter.net/iau/MPCORB/MPCORB.DAT.}}. 

The six regions are defined as follows: 

\begin{enumerate}
\item Inner belt: 2.1 AU < $a$ < 2.5 AU (3:1J);
\item Middle belt: 2.5 AU < $a$ < 2.823 AU (5:2J);
\item Pristine belt: 2.823 AU < $a$ < 2.956 AU (7:3J); 
\item Outer belt: 2.956 AU < $a$ < 3.28 AU (2:1J);
\item  Cybele belt: 3.28 AU < $a$ < 3.51 AU;
\item  High-Inclination belt:  $\sin i $ > 0.34 ($i > 20 \degr$) ( $\nu_{6}$ secular resonance).
\end{enumerate}

\begin{figure}[htp]
\centering
\includegraphics[width=8cm]{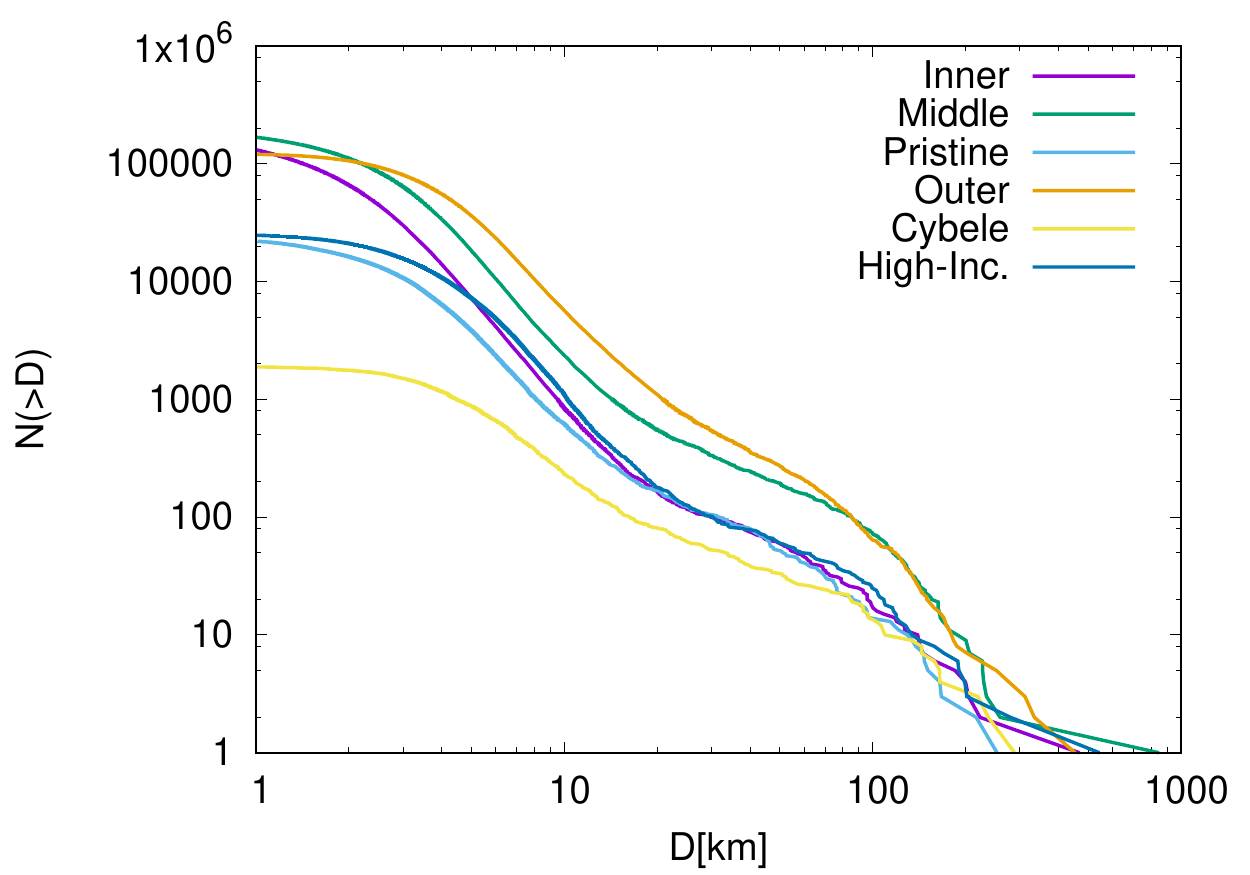}
\caption{Observed SFDs $N\left(>D\right)$ of the six regions of the MB \citep{Cibulkova2014}.}
    \label{fig:MB6}
\end{figure}

The observed SFDs of the different regions of the MB are plotted in Fig. \ref{fig:MB6}, which were constructed by \cite{Cibulkova2014}. These latter authors built the observed SFDs by calculating asteroid sizes using data of asteroids with known albedos in the AstOrb catalog \citep[by IRAS observations,][]{Tedesco2002} and from the WISE satellite \citep{Masiero2011}. For those asteroids in the AstOrb catalog  which have no albedo measurements, \cite{Cibulkova2014} calculated the diameter from H-magnitude \citep{Bowell1989} by randomly assigning albedos from a distribution obtained from WISE data. They carefully verify the validity of the method.
We see that the Middle and Outer are the most populated regions, followed by Inner and High-Inclination belts, while the Pristine and Cybele belts have the smallest number of asteroids. 

To model the collisional evolution of the MB, it is necessary to know  the intrinsic probabilities of collisions  $P_{\textrm{imp}}$ and mutual impact velocities $v_{\textrm{imp}}$ between the asteroids of the different regions. These values were calculated by \cite{Cibulkova2014} using the code written by \cite{BottkeGreenberg1993}. The mean values are listed in Table \ref{tab:ProbVimp}. 

\begin{table}
\caption{Intrinsic collisional probabilities $P_{\textrm{imp}}$ and mutual impact velocities $v_{\textrm{imp}}$ between bodies belonging to the different regions of the MB \citep{Cibulkova2014}.} 
\begin{center}
\begin{tabular}{|c c c|}
\hline
\hline
Populations & $P_{\textrm{imp}} \left(10^{-18} \text{km}^{-2} \text{yr}^{-1}\right)$ & $v_{\textrm{imp}} \left(\text{km }  \text{s}^{-1}\right)$ \\
\hline
\hline
Inner-Inner    & 11.98 & 4.34 \\
Inner-Middle   & 5.35  & 4.97 \\
Inner-Pristine & 2.70  & 3.81 \\
Inner-Outer    & 1.38  & 4.66 \\
Inner-Cybele   & 0.35  & 6.77 \\
Inner-High Inc.& 2.93  & 9.55 \\
\hline
Middle-Middle   & 4.91  & 5.18 \\
Middle-Pristine & 4.67  & 3.96 \\
Middle-Outer    & 2.88  & 4.73 \\
Middle-Cybele   & 1.04  & 5.33 \\
Middle-High Inc.& 2.68  & 8.84 \\
\hline
Pristine-Pristine & 8.97  & 2.22 \\
Pristine-Outer    & 4.80  & 3.59 \\
Pristine-Cybele   & 1.37  & 4.57 \\
Pristine-High Inc.& 2.45  & 7.93 \\
\hline
Outer-Outer    & 3.57  & 4.34 \\
Outer-Cybele   & 2.27  & 4.45 \\
Outer-High Inc.& 1.81  & 8.04 \\
\hline
Cybele-Cybele   & 2.58  & 4.39 \\
Cybele-High Inc.& 0.98  & 7.87 \\
\hline
High Inc.-High Inc.& 2.92  & 10.09 \\
\hline

\end{tabular}
\end{center}

\label{tab:ProbVimp}
\end{table}

\section{The model}
We built a \texttt{FORTRAN} code,  \texttt{ACDC,}  based on the prescriptions of the \texttt{BOULDER} code developed by \cite{Morbidelli2009} and the adaptation to six populations made by \cite{Cibulkova2014}. 

\texttt{ACDC} is a multi-population code that simulates the collisional and dynamical evolution of the MB by evolving in time the incremental number of bodies of size $D$ ($N(D,t)$) in each one of the defined regions. Such evolution, summarized in Eq. \ref{eq:N(t)}, is determined by the change in the number of bodies due to the objects being destroyed and fragments being ejected in collisions ($\Delta N_{\text{COL}}\left(D,t\right)$), which can be positive or negative, and by the dynamical depletion of the MB by the Yarkovsky effect ($ \Delta N_{\text{YE}}\left(D,t\right)$), which we assume to be negative following the approach of \cite{Bottke2005a}, \cite{OBrien2005}, and \cite{deElia2007}. \texttt{ACDC} is a statistical code, as it involves a random number generator and Poisson statistics, and so different seeds produce different results. 
\begin{equation}
N\left(D,t+\Delta t\right)=N\left(D,t\right)+\Delta N_{\text{COL}}\left(D,t\right) + \Delta N_{\text{YE}}\left(D,t\right).
\label{eq:N(t)}
\end{equation}
The \texttt{ACDC} code uses a set of fixed discrete logarithmic size bins. In particular, we used 135 size bins  for this work, with central values in the range $1  \text{ mm}\leq D\leq 980  \text{ km}$ in diameter in such a way that from one bin to the next the diameter increases by a factor of $10^{1/15}$. 

In this section, we describe the model and all the methods and algorithms implemented in this work.

\subsection{Asteroid disruptions - Scaling law}
The scaling law is a key issue in the model of asteroid collisions, as it dictates the outcome of the impact event. It is determined by the specific energy $Q^{*}_{D}$, which is defined as the energy per unit mass required to catastrophically disrupt an asteroid (i.e., such that half of the mass is dispersed). 

 \cite{BenzAsphaug} derived a functional form for the scaling law using SPH simulations for basaltic and icy targets in different impact speeds, given by:
\begin{equation}
Q_{\text{D}}^{*}=Q_{0}R^{a}+B\rho R^{b},
\label{eq:QD}
\end{equation}
where $R$ is the target radius in centimeters (cm), $\rho$ the target density in $\text{g cm}^{-3}$, $Q_{0}$ and $B$ normalization parameters, and $a$ and $b$ the slopes of the corresponding power law.

The scaling law that has been used so far in most collisional evolution works of the MB  is the one derived by \cite{BenzAsphaug} for monolithic basaltic targets at  5 $\text{km s}^{-1}$ impact speeds. However, using the same law for all the bodies in the asteroid belt implies the assumption that all bodies have the same composition and collide with the same velocities. Previous works attempted to obtain scaling laws for different materials. For example, \cite{Cibulkova2014} constructed new $Q^{*}_{\text{D}}$ functions for rubble piles using the results of \cite{Benavidez2012}, who ran a set of SPH simulations of disruption of $D$ = 100 km rubble-pile targets. However, \cite{Cibulkova2014} found that monolithic asteroids provide a better match with the observed data than rubble-piles. \cite{Bottke2005a} and \cite{Cibulkova2014} also tested many different scaling laws by changing the exponent $b$ in Eq. \ref{eq:QD}, and concluded that laws much different from that of \cite{BenzAsphaug} cannot be used for the MB because they fail to reproduce the observed asteroid families.

In our simulations, we therefore decided to use the same $Q^{*}_{\text{D}}$ scaling law for all the objects in the MB, namely that derived by \cite{BenzAsphaug} for basaltic materials at 5 $\text{km s}^{-1}$ impact speeds. The values of the parameters involved in the calculation of $Q_{\text{D}}^{*}$ are listed in Table \ref{tab:Benz}. 

\begin{table}
\caption{Parameters of the scaling law for monolithic basaltic targets and impact speeds of $5$ km s$^{-1}$, derived by \cite{BenzAsphaug} . $\rho$ is the target density in g cm$^{-3}$, $Q_{0}$ and $B$ normalization parameters, and $a$ and $b$ the slopes of the corresponding power law.} 
\begin{center}
\begin{tabular}{|c c c c c|}
\hline
\hline
$\rho\left(\text{g cm}^{-3}\right)$ & $Q_{0} \left(\text{erg g}^{-1}\right)$ & $a$ & $B (\text{erg g}^{-1})$ & $b$ \\
\hline
\hline
3.0 & $9\times10^{7}$ & -0.36 & 0.5 & 1.36 \\
\hline

\end{tabular}
\end{center}

\label{tab:Benz}
\end{table}

\subsection{Outcome of a single collision}
Here, we present the model used to describe the outcome of a single collision between two bodies, which is based on the prescriptions of the \texttt{BOULDER} code \citep{Morbidelli2009}.

We name the bigger body the target, or the parent body, with mass $m_{i}$, and the smaller body the projectile, or impactor, with mass $m_{j}$. We assume that all bodies are spherical with uniform density $\rho$.

After an impact event, fragments are ejected into space. The largest of these fragments has a mass $M_{\text{LF}}$. The SFD of the fragments is represented by a cumulative power law, given by:
\begin{equation}
F\left(\geq D\right)=AD^{q},
\label{eq:Nfrag}
\end{equation}
where $D$ is the fragment diameter, $q$ a given slope, and $A$ an appropriate normalization constant, considering that there is only one largest fragment after the collision.
The largest surviving body after the collision is called the largest remnant, with mass $M_{\text{LR}}$. 

The kinetic energy of the impact is the fundamental quantity that determines the outcome of the collision. In particular, we use the specific impact energy of the projectile $Q$, given by:
\begin{equation}Q=\frac{1}{2}\frac{m_{j} v_{\text{imp}}^{2}}{\left(m_{i}+m_{j}\right)},\end{equation}where $v_{\text{imp}}$ is the impact velocity. The values of $v_{\text{imp}}$ are listed in Table \ref{tab:ProbVimp}.

 We compare the specific impact energy with the scaling law $Q_{\text{D}}^{*}$ (Eq. \ref{eq:QD}) and consider two impact regimes: 
\begin{itemize}
\item If $Q<Q_{\text{D}}^{*}$ , we have a cratering event. The target ejects fragments into space and a crater is created in the  surface of the target. 
\item If $Q \geq Q_{\text{D}}^{*}$, we have a catastrophic disruption event, in which more than half of the total initial mass is dispersed in the form of fragments. 
\end{itemize}

The expressions for $M_{\text{LR}}$ and $M_{\text{LF}}$ derived by \cite{BenzAsphaug} and \cite{Durda2007} are:
\begin{equation}
M_{\text{LR}}=\begin{cases}
\left[-\frac{1}{2}\left(\frac{Q}{Q_{\text{D}}^{*}}-1\right)+\frac{1}{2}\right]\left(m_{i}+m_{j}\right), & Q<Q_{\text{D}}^{*} \text{ (cratering)}\\
\\
\left[-\frac{7}{20}\left(\frac{Q}{Q_{\text{D}}^{*}}-1\right)+\frac{1}{2}\right]\left(m_{i}+m_{j}\right), & Q \geq Q_{\text{D}}^{*} \text{ (catastrophic)}
\end{cases}
 \label{eq:MLR}
\end{equation}
and
\begin{equation}
M_{\text{LF}}=8\times10^{-3}\left[\frac{Q}{Q_{\text{D}}^{*}}\exp\left(-\left(\frac{Q}{4Q_{\text{D}}^{*}}\right)^{2}\right)\right]\left(m_{i}+m_{j}\right).
\end{equation}
If $M_{\text{LR}}$ is negative, we assume that the target is pulverized and all mass is lost below the smallest bin.

The slope of the SFD of the fragments is given by:
\begin{equation}
q=-10+7\left(\frac{Q}{Q_{\text{D}}^{*}}\right)^{0.4}\exp\left(-\frac{Q}{7Q_{\textrm{D}}^{*}}\right).
\end{equation}
Since such a steep slope can lead to an unrealistic infinite mass, we assume that the fragment size distribution has a cumulative slope $q=-2.5$ \citep{dohnanyi1969} below a turn-over diameter $D_{\textrm{t}}$. We compute $D_{\textrm{t}}$ so that the mass integral of the fragments is equal to the total ejected mass.

\subsection{Collisional evolution}
\label{sec:ColEVol}
Here, we describe the method used to simulate the collisional evolution of the MB, which consists in modifying the incremental number of objects in every size bin during each timestep. The total integration time is 4000 Myr. 

We split the MB in six regions of index $r$, where $r=1,...,6$. $i$ and $j$ denote the target and projectile indexes, with diameters $D_{i}$ and $D_{j}$ and belonging to populations $r_{i}$ and $r_{j}$, respectively. 

Before the simulation starts, we arrange two quantities for all possible target--projectile pairs: the frequency of each possible collision and their outcomes. The first is given by: \begin{equation}f_{i,j}^{r_{i},r_{j}}=\frac{1}{4}P_{r_{i},r_{j}}\left(D_{i}+D_{j}\right)^{2},
\label{eq:Frec}
\end{equation}
where the values of the intrinsic impact probabilities $P_{r_{i},r_{j}}$  are listed in Table \ref{tab:ProbVimp} \citep{Cibulkova2014}.  

Following the algorithm outlined by \cite{OBrien2005}, we generate 3D arrays $F_{i,j,k}^{r_{i},r_{j}}$, where $k$  is the size bin. These arrays give the incremental fragment SFD resulting from every possible collision derived from Eq.\ref{eq:Nfrag}. The $F$ arrays are symmetrical between targets and projectiles, and hence between regions, since the collision probabilities and impact velocities are the same regardless of which body is the target or projectile. We create one array for every pair of regions and the number of  $F$ arrays created in this six-region model is 21. 

At this point we begin the simulation. First, we calculate the integration time step $\Delta t$. To do so, we calculate the collisional lifetimes $\tau_{i}^{r_{i}}$ of all asteroids from all regions, by summing all the projectiles that can catastrophically disrupt them:
\begin{equation}
\tau_{i}^{r_{i}}=\left(\sum_{r_{j}=1}^{6}\sum_{D_{j}=D^{i}_{\text{disrupt}}}^{D_{i}}N_{j}^{r_{j}}f_{i,j}^{r_{i},r_{j}}\right)^{-1}, 
\end{equation}
where $N_{j}^{r_{j}}$ is the incremental number of projectiles and $D_{\text{disrupt}}^{i}$ the diameter of the smallest projectile that can catastrophically disrupt the target. 

The timestep $\Delta t$ is therefore calculated as ten times smaller than the minimum $\tau_{i}^{r_{i}}$ value. By doing so, we assure that, from one timestep to the other, the removal of bodies in a size bin never exceeds 10\%. 

The change in the number of bodies of a given size is determined by two quantities: the number of collisions, and the outcome of each event (i.e., bodies destroyed and fragments created). 

We calculate the deterministic number of collisions between all  target--projectile  pairs in the six regions of the MB:
\begin{equation}
nc_{i,j}^{r_{i},r_{j}}=f_{i,j}^{r_{i},r_{j}}N_{\text{pairs}}^{r_{i},r_{j}}\Delta t,
\label{eq:ncol}
\end{equation}
where $N_{\textrm{pairs}}^{r_{i},r_{j}}$ is the number of target-projectile pairs, which is determined by $N^{r_{i}}_{i}$ and $N^{r_{j}}_{j}$, the number of targets and projectiles of diameter $D_{i}$ and $D_{j}$, respectively. If both target and projectile belong to the same size bin and population, their numbers are equal, and so the number of pairs is given by:
\begin{equation}
N_{\text{pairs}}^{r_{i},r_{j}}=\frac{N_{i}^{r_{i}}\left(N_{j}^{r_{j}}-\delta_{i,j}\delta_{r_{i},r_{j}}\right)}{1+\delta_{i,j}\delta_{r_{i},r_{j}}},
\end{equation}
where the $\delta$ function is the Kronecker delta. 

If any number of collisions between a given target--projectile pair is smaller than a limit value of $5$, we recalculate it with Poisson statistics using the result given by Eq. \ref{eq:ncol} as the mean value and a random seed \citep{Press1992}. Therefore, in a given timestep, the number of collisions can be an integer number between 0 and 5. 

Next, we calculate the change in the number of bodies of a given size bin $k$ in a region $r $ by taking into account all of the bodies destroyed as well as the new fragments created: 
\begin{equation}
\Delta {N_{\text{COL}}}_{k}^{r_{i}}=\sum_{r_{j}=1}^{6}\Delta n_{k}^{r_{i},r_{j}}.
\end{equation}
If both target and projectile belong to the same region, the change is simply given by the number of collisions multiplied by the number of fragments created in each bin; however, if they belong to different regions, mixing between targets and projectiles from the distinct regions must be considered by removing projectiles from one region and adding them to the other. The expressions of the interaction terms, which were adapted from \cite{OBrien2005}, are given by:
\begin{equation}
\Delta n_{k}^{r_{i},r_{j}}=\sum_{i=1}^{n}\sum_{j=1}^{i}\begin{cases}
\vspace{5mm}
nc_{i,j}^{r_{i},r_{i}}F_{i,j,k}^{r_{i},r_{i}}, & r_{i} = r_{j}\\
\vspace{5mm}
\left(F_{i,j,k}^{r_{i},r_{j}}+\delta_{j,k}\right)nc_{i,j}^{r_{i},r_{j}}\gamma_{i,j}-\delta_{jk}nc_{i,j}^{r_{j},r_{i}}, & r_{j}>r_{i}\\
\vspace{5mm}
\left(F_{i,j,k}^{r_{j},r_{i}}+\delta_{j,k}\right)nc_{i,j}^{r_{j},r_{i}}-\delta_{j,k}nc_{i,j}^{r_{i},r_{j}}\gamma_{i,j}, & r_{j}<r_{i}
\end{cases}
,\end{equation}
where we name $\gamma_{i,j}=1-\delta_{i,j}$ and $n$ is the number of size bins. 

Since our model deals with a finite number of bins, the bodies able to disrupt the smallest bodies are generally smaller than the first bin. Ignoring the disruption of the smallest bodies may lead to unrealistic waves in the size distribution \citep{CampoBagatin1994,DurdaDernott1997}. In order to prevent these undesirable artificial waves, we do not calculate the collisional evolution of the bodies smaller than 10 cm; instead, we extrapolate them from the larger bodies after each timestep.

\subsection{Dynamical evolution: Yarkovsky effect}

The Yarkovsky effect is a radiation force that produces the change in the orbital parameters of small rotating asteroids because of the asymmetry between the direction of absorption of sunlight and the direction of re-emission of thermal radiation. The radiated energy causes a force along the body's orbit that in turn causes asteroids smaller than tens of kilometers to slowly drift in semi-major axis, allowing them to reach resonances with the planets that drive them into planet-crossing orbits. Thus, the Yarkovsky effect causes a dynamical depletion of the MB, and can be considered a ``sink'' for small asteroids. This depletion affects the collisional evolution of the MB, because fewer smaller bodies means fewer collisions with larger bodies, and therefore it has to be considered in any collisional evolution model. 

The general expressions for the diurnal and seasonal Yarkovsky effects are \citep{Peterson1976,Burns1979,Rubincam1995,Vokrouhlicky1998,Farinella1998b,Vokrouhlicky1999,ReviewYarko2015}:
\begin{equation}
\dot a_{\text{d}}=-\frac{8}{9}\frac{\alpha\Phi}{n}W\left(R_{\omega},\Theta_{\omega}\right)\cos\gamma,
\label{eq:YarkoD}
\end{equation}
\begin{equation}
\dot a_{\text{s}}=\frac{4}{9}\frac{\alpha\Phi}{n}W\left(R_{n},\Theta_{n}\right)\sin^{2}\gamma,
\label{eq:YarkoS}
\end{equation}
where $\alpha=1-A_{\textrm{B}}$, $A_{\textrm{B}}$ denoting the Bond albedo \citep{Vokrouhlicky2001}, $\Phi=\pi R^{2}F/\left(mc\right)$, with $R$ being the radius of the body, $F$ the solar radiation flux at the orbital distance $a$ from the Sun, $m$ the mass of the body, $c$ the speed of light, $n$ the orbital mean motion, $\omega$ the rotation frequency, and $\gamma$ the spin-axis obliquity. 

The function $W\left(R_{\nu},\Theta_{\nu}\right)$ depends on parameters like the thermal conductivity $K$, the thermal capacity $C$, the density $\rho$, and a frequency $\nu$, which is equal to $n$ for the seasonal effect and $\omega$ for the diurnal effect. It is determined by the radius of the body $R$, which is scaled by the penetration depth $l_{\nu}=\sqrt{K/ \rho C\nu}$ of the thermal wave, and the thermal parameter $\Theta_{\nu}=\sqrt{K\rho C} \sqrt{\nu}/\left(\epsilon \sigma T^{3}\right)$. 
 The function $W$ is given by:
\begin{equation}W\left(R_{\nu},\Theta_{\nu}\right)=-0.5\frac{\Theta_{\nu}}{1+\Theta_{\nu}+0.5\Theta_{\nu}^{2}}.
\end{equation}
The expression for the seasonal effect (Eq. \ref{eq:YarkoS}) is accurate for bodies bigger than $4l_{\nu}$ \citep{Farinella1998b}. We assume that for bodies smaller than $4 l_{\nu}$, $\dot a_{\text{s}}\propto D^{3/2}$  \citep{OBrien2005}.

In the thermal model, following \cite{Cibulkova2014}, we assume a break in the thermal conductivity that reflects the rotational parameters of small bodies \citep{Warner2009}: $K = 0.01 \text{ W} \text{ m}^{-1} \text{ K}^{-1}$ for $D>D_{\textrm{YE}}$ and $K = 0.01 \text{W} \text{ m}^{-1} \text{ K}^{-1}$ for $D \leq D_{\textrm{YE}}$. We assume the value of the transition value $D_{\textrm{YE}}$ = 200 m and a size-dependent spin rate $\omega(D)=\frac{2\pi}{P_{0}}\frac{D_{0}}{D}$, $P_{0}$ = 5 h, $D_{0}$ = 5 km. The thermal capacity assumed is $C = 680  \text{ J}  \text{ kg}^{-1} \text{ K}^{-1} $, the infrared emissivity $\epsilon = 0.95$, and the Bond albedo $A_{\textrm{B}} = 0.02$. As the diurnal and seasonal effects depend on $\cos\gamma$ and $\sin \gamma$, respectively, we consider a mean obliquity of $45\degr$ in order to calculate a mean Yarkovsky effect. The physical densities considered for the bodies of the different regions of the MB are shown in Table \ref{tab:Yark}.

\begin{figure}[htp]
    \centering
\includegraphics[width=8cm]{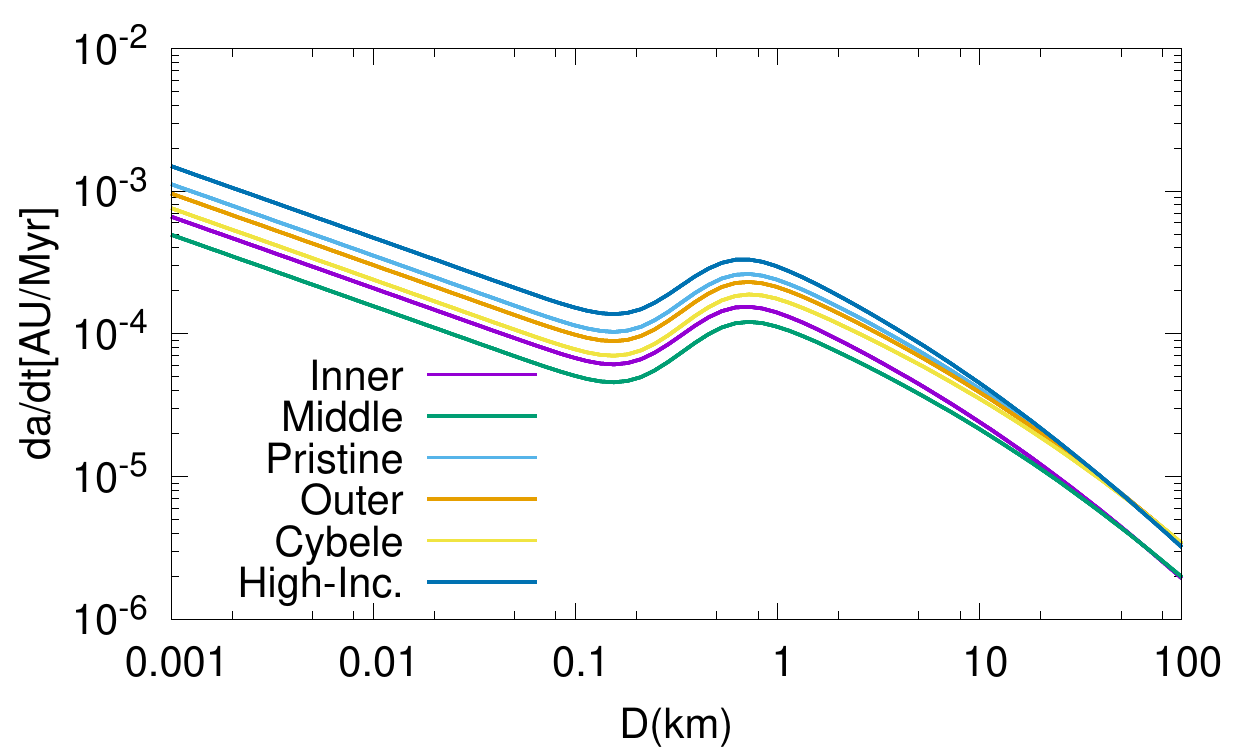}
\includegraphics[width=8cm]{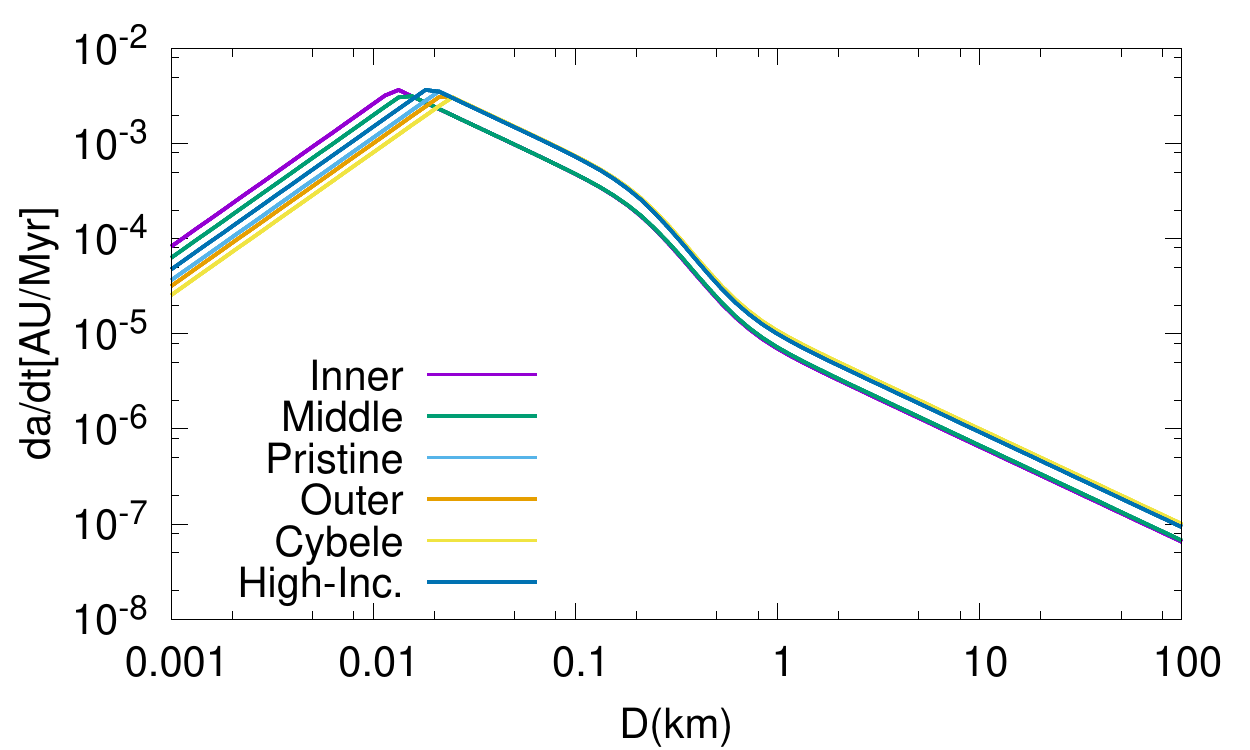}
\includegraphics[width=8cm]{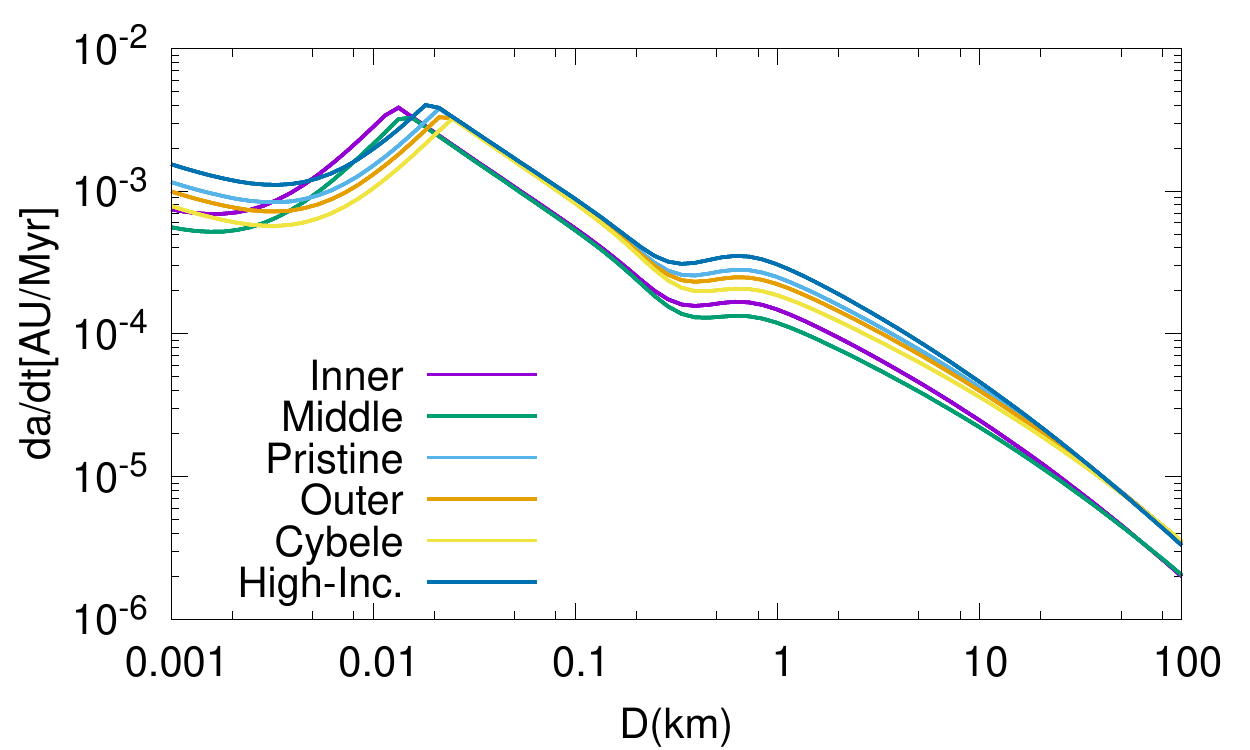}
    \caption{Semimajor-axis drift rate $\dot a$ due to the Diurnal (top), Seasonal (middle), and total (Bottom) Yarkovsky effect as a function of size, for the six regions of the MB. A color version of this figure is available in the electronic version of this article.}
    \label{fig:dadt}
\end{figure}

\begin{figure}[htp]
    \centering
\includegraphics[width=8cm]{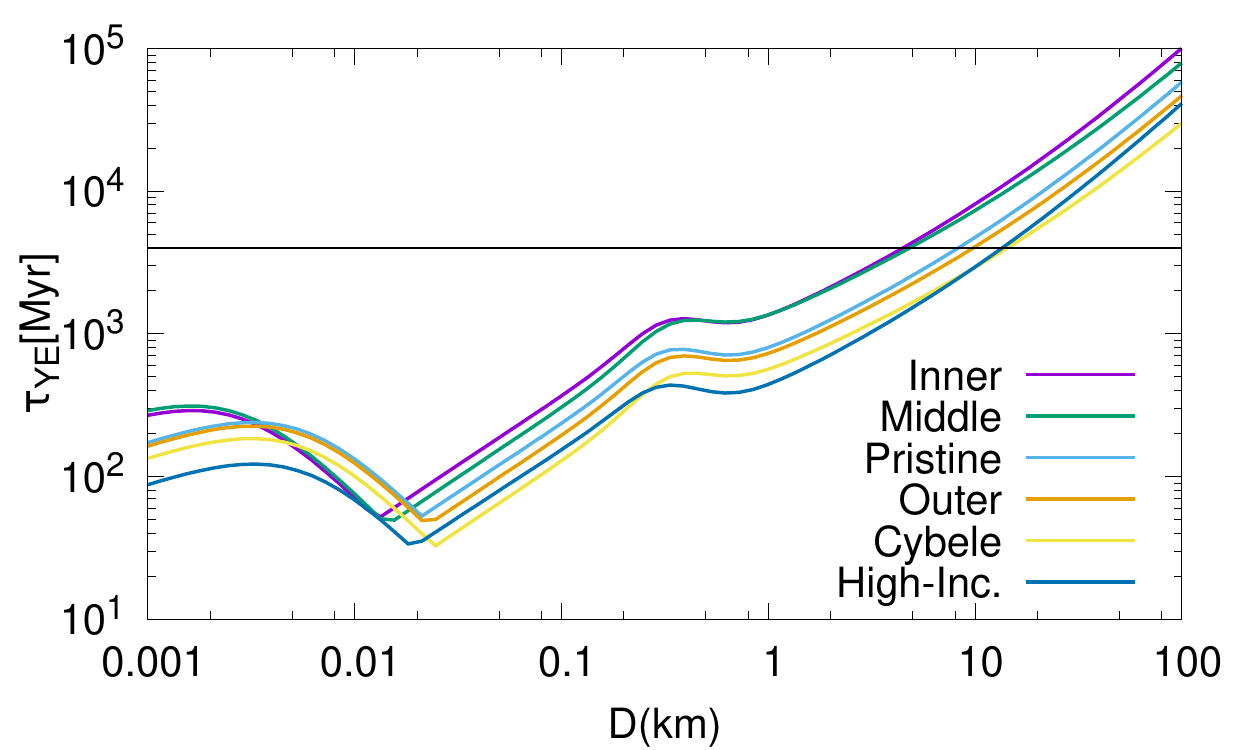}
    \caption{ Timescale $\tau_{\textrm{YE}}$ of the Yarkovsky effect as a function of size $D$ for the six regions of the MB. The black line indicates the total integration time of 4000 Myr. A color version of this figure is available in the electronic version of this article.}
    \label{fig:TYE}
\end{figure}

Following \cite{Cibulkova2014}, we assume that the Yarkovsky effect causes an exponential decay described by:
\begin{equation}
N_{\textrm{YE}}^{r}\left(D,t+\Delta t\right)=N^{r}\left(D,t\right)\exp\left(-\frac{\Delta t}{\tau_{\text{YE}}^{r}\left(D\right)}\right),
\end{equation}
\noindent{where} $N_{k}^{r}\left(t\right)$ is the number of bodies of diameter $D$ from region $r$ at time $t$, and $\Delta t$ is the timestep. In particular, $\tau_{\text{YE}}^{r}$ is the characteristic timescale of the Yarkovsky effect, which is defined as the mean time it takes for asteroids to reach a resonance: 
\begin{equation}
\tau_{\text{YE}}^{r}\left(D\right)=\frac{\Delta a_{r}}{\dot{a}_{r}},
\label{eq:TYE}
\end{equation}
where $\Delta a_{r}$ is half of the region size in semi-major axis (see Table \ref{tab:Yark}), and $\dot{a}_{r}$ is the sum of the rates associated with the diurnal and seasonal Yarkovsky effects calculated with Eqs. \ref{eq:YarkoD} and \ref{eq:YarkoS}, respectively. The values of $\dot{a}$ as a function of size, for the diurnal, seasonal, and total Yarkovsky effect are plotted for each region in Fig. \ref{fig:dadt}. We find a total drift rate of $1-3\times 10^{-4} \text{AU Myr}^{-1}$ for 1 km bodies, a range that contains typical values of $2\times 10^{-4} \text{AU Myr}^{-1}$ \citep{Granvik2017}. The timescales for the six regions are plotted in Fig. \ref{fig:TYE}, which are consistent with the ones derived by \cite{Cibulkova2014}.

In summary, the total number of bodies removed from a region due to the Yarkovsky effect to be included in Eq. \ref{eq:N(t)} is $\Delta N_{\textrm{YE}}^{r} = N_{\textrm{YE}}^{r}\left(D,t+\Delta t\right) - N^{r}\left(D,t\right)$. It is important to note that this general dynamical model, formulated as it is, assumes two things: when asteroids fall into resonances, they are immediately removed from the MB and added to the NEA region, and all resonances remove asteroids with the same efficiency. However, since the timescale depends on the size of the region ($\Delta a$ in Eq. \ref{eq:TYE}), we observed that the timescale for the Pristine region is very short, and causes an excessive depletion in this region, which is narrower and less populated than the other regions. \cite{Gladman} studied the dynamical lifetimes of asteroids injected into the resonances and calculated the time it takes to cross the orbit of Earth. These latter authors found times shorter than 1 Myr for strong resonances, like $\nu_{6}$, 3:1J and 5:2J. However, they found that if an asteroid falls into the 7:3J resonance, which is the one that separates the Pristine from the Outer belt, it takes $\sim$6.3 Myr to cross Earth's orbit. Taking this particular feature of the Pristine belt into consideration in our general Yarkovsky model, we found that we can counter the excessive dynamical depletion and make a better fit with observations if we multiply the timescale of the Pristine region by a factor of three. By doing this, we enable the asteroids that fall into the 7:3J resonance to stay in the Pristine region for a longer time before being removed from the asteroid belt. It is worth noting that our solution found for Pristine takes it account the many uncertainties involved in a general Yarkovsky model, as well as the particular characteristics of that region.

\begin{table}[h]
\caption{Parameters of the Yarkovsky effect for the different regions of the MB \citep{Cibulkova2014}. $\Delta a$ is half of the region size or typical distance from neighboring resonances, and $\rho$ is the bulk and surface density assumed for the respective bodies. } 
\begin{center}
\begin{tabular}{|c c c |}
\hline
\hline
Region & $\Delta a$ (AU) &  $ \rho\left(\text{kg m}^{-3}\right)$ \\
\hline
\hline
Inner & 0.2 & 2500 \\
Middle & 0.1615 & 2500 \\
Pristine & 0.0665 & 1300 \\
Outer & 0.162 & 1300 \\
Cybele & 0.105 & 1300 \\
High-Inc & 0.135 & 1300 \\
\hline

\end{tabular}
\end{center}

\label{tab:Yark}
\end{table}

\subsection{Fit with observational data}

As mentioned above, the code developed in this work is a statistical model that treats big impacts as random events using Poisson statistics. Therefore, runs using different seeds for the random number generator produce different results, and so we develop a set of runs using different random seeds and then interpret the results statistically. 

To obtain a quantitative measure of how well a simulation reproduces observational data, we follow the procedure described by \cite{Bottke2005b} and \cite{Cibulkova2014}. The metric used to determine the goodness of fit between the observed size distribution of a region ($N_{\text{obs}}$) and the simulation results ($N_{\text{sim}}$) is a hybrid $\psi^{2}$ test:
\begin{equation}
\psi^{2}=\sum_{i}\left(\frac{N_{\text{sim}}\left(>D_{i}\right)-N_{\text{obs}}\left(>D_{i}\right)}{\sigma_{i}}\right)^{2},
\end{equation}
where the summation extends over a range of diameters $D_{\text{min}}$ ($\sim$ 3 -- 6 km) and $D_{\text{max}}\sim 250$ km, estimated from the observed SFDs \citep{Cibulkova2014}. The uncertainties are given by $\sigma_{i}=10\% N_{\text{obs}}\left(>D_{i}\right)$. For each run, we calculate the individual metrics $\psi^{2}_{r}$  for the six regions and then a global metric $\psi^{2}_{\text{MB}}$ for the MB as a whole, and also a mean metric $\psi^{2}_{\text{MEAN}}$, by simply averaging the metrics of the six regions. 

The ideal case is to find a set of runs that produce good fits in the individual regions and globally. However, this is very unlikely to happen in a single run due to the high stochasticity and randomness of big impact events and the mutual interactions between the different regions of the MB. It is possible that some runs may produce good fits in some populations and simultaneously bad fits in other populations. Also, some runs may produce bad fits in individual populations, but produce a good match with the global MB. Moreover, some runs can even produce impossible results like Vesta colliding with Ceres and thus destroying the most massive bodies in the asteroid belt. Therefore,  a very low number of acceptable runs is expected. The criterion adopted in this work is to select runs by sorting the ones with the lowest $\psi^{2}_{\text{MEAN}}$ metrics and then calculate medians of the desired results. 

\section{Dynamical evolution of the NEAs}
\label{Sect:neas}
In this section, we present our modeling of the dynamical evolution of the NEAs. The asteroids that escape from the MB due to the combined action of the Yarkovsky effect and resonances are delivered to the NEA population \citep[e.g.,][]{Gladman,Bottke2002,Granvik2016}. The asteroids in Earth-crossing orbits reside in that region for some time until they reach a sink. The most important sinks of NEAs are collisions with the Sun, escape from the inner Solar System after a close encounter with Jupiter \citep{Granvik2018}, destruction by thermal effects \citep{Granvik2016}, and collisions with terrestrial planets. It was shown by \cite{Bottke2002} and \cite{Granvik2018} that the different entrance routes yield different mean lifetimes. The different regions of the MB that we take into account in this work are bounded by the resonances that \cite{Granvik2018} consider as entrance routes. Therefore, a single region of the MB can deliver asteroids to the NEA population through different entrance routes. 

Based on our partition of the MB (see Fig. \ref{fig:MB}), we consider the $\nu_{6}$ and 3:1J resonances as the entrance routes for the NEAs coming from the Inner belt,  3:1J and 5:2J for those coming  from the Middle belt, 5:2J and 2:1J for those coming  from the Pristine and Outer belts,  2:1J for those coming from the Cybele belt, and 3:1J, 5:2J, 2:1J, and Phocaeas for those coming from High-Inclination belt \citep{Granvik2017}.
 
In order to quantify the contribution of the different regions of the MB to the NEAs, we calculate the mean dynamical lifetimes of NEAs according to their source regions. To do so, for each of our regions we average the lifetimes of bodies $\tau$, weighted by the contribution of NEAs coming from each of the surrounding resonances. In particular, we use the values for the average lifetimes $L$ in the NEA region and the prediction for the debiased number of NEAs coming from each of the resonances derived by the dynamical simulations performed by \cite{Granvik2018}, listed in Table \ref{tab:Tnea1}. Thus, we calculate the mean lifetimes of NEAs coming from the given regions as follows:

\begin{equation}
\tau_{r} = \frac{\sum L_{i} \alpha_{i} h_{i} N_{i}}{\sum \alpha_{i}h_{i} N_{i}}
\label{eq:TauNea}
.\end{equation}
We split the number of NEAs $N$ that are delivered through a resonance according to the contribution from each region. The values $\alpha$ in Eq. \ref{eq:TauNea} represent, for each resonance, the fraction of $N$ that comes from the adjacent regions. For example, we consider $\alpha=0.5$ for 3:1J, which means that half of the bodies that enter in the NEA region via the 3:1J resonance are delivered by the Inner belt while the other half is delivered by the Middle belt, as was proposed by \cite{Bottke2002}. Similarly, as our High-Inclination belt contains all asteroids in the MB with inclinations greater than 20$\degr$, we split $N$ according to $h$, the fraction of NEAs that enter a resonance from a low or high inclination region. For example, we assume $h=0.8$  for 3:1J, which means that 80\% of bodies come from inclinations smaller than 20\degr while 20\% come from inclinations greater than 20\degr. We estimated these fractions with the simulations performed by \cite{Granvik2017}. The values we assumed for the $\alpha$ and $h$ fractions are listed in Table \ref{tab:Tnea2}. 

The Pristine and Outer belt are limited by 5:2J, 7:3J, and 2:1J, but the effect of 7:3J is not clear  from the dynamical simulations of \cite{Granvik2017,Granvik2018} . Moreover, it shows that the asteroids from the Outer belt are able to cross the 7:3J resonance and enter the Pristine region, a process we are not considering in this model. Thus, we assign the same time to the Outer, Pristine, and Cybele regions by averaging the times of 5:2J and 2:1J, with Eq. \ref{eq:TauNea}.

The High-Inclination region in our collisional model spans the entire MB in semi-major axis, and therefore passes through all the major resonances present there, each having different associated lifetimes as Table \ref{tab:Tnea1} shows. Therefore, we split the contribution of the High-Inclination from our model into three different subregions corresponding to the Inner, Middle, and Outer ranges in semi-major axis. We assign them 80\%, 5\%, and 15\% of the High-Inclination contribution, respectively, consistent with the outputs of the simulations performed by \cite{Granvik2017}, and calculate their dynamical times separately with Eq. \ref{eq:TauNea}. 

Table 5 summarizes the mean dynamical times for NEAs coming from the different regions of the MB, calculated with Eq. \ref{eq:TauNea}, as well as the way we group the contribution for each resonance.

\begin{table}[h]
\begin{center}
\caption{The prediction for the debiased number of NEAs $N$ coming from each entrance route in the range $17<H<25$, and the average lifetime in the NEA region $L$ \citep{Granvik2018}. } 
\label{tab:Tnea1}
\begin{tabular}{|c c c |}
\hline
\hline
 Entrance route & $N$ & $L$ [Myr] \\
\hline
\hline
$\nu_{6}$ & 296\,400 &  6.91 \\
3:1 & 286\,400 & 1.86 \\
5:2 & 7\,200    &       0.68 \\
2:1& 4\,400     &       0.40 \\
Phocaeas & 1\,300       &       11.16 \\
\hline 
\hline 
\end{tabular}
\end{center}
\end{table}

\begin{table}[h]
\begin{center}
\caption{Mean dynamical lifetimes of NEAs coming from the different regions of the MB, calculated according to the parameters listed in Table \ref{tab:Tnea1} \citep{Granvik2018}, the simulations performed by \cite{Granvik2017}, and Eq. \ref{eq:TauNea}. Here, $\alpha$ is the fraction of asteroids from the bounding resonance that come from the corresponding region, while $h$ is the fraction that comes from the high or low inclination region. } 
\label{tab:Tnea2}
\begin{tabular}{|c c c c c |}
\hline
\hline
 Region & Entrance route & $\alpha $& $h$ & $\tau$ [Myr] \\
\hline
\hline
Inner  & $\nu_{6}$ & 1 & 1 & 5.54 \\
       & 3:1J & 0.5 & 0.8 & \\
\hline
Middle & 3:1J & 0.5 & 0.8 & 1.79 \\
                & 5:2J  & 0.5 & 0.9   & \\
\hline
Outer,Pristine  & 5:2J &0.5 & 0.9 & 0.53 \\
and Cybele      & 2:1J & 1 &0.8 & \\
\hline
High-Inc:Inner & Phocaeas & 1 & 1 & 2.04 \\                      
                                & 3:1J & 0.95 & 0.2 & \\
\hline
High-Inc: Middle & 3:1J & 0.05 & 0.2 & 1.81   \\
                                                & 5:2J & 0.95 & 0.1 & \\
\hline 
High-Inc: Outer & 5:2J & 0.05 & 0.1 & 0.52 \\
                          & 2:1J & 1    & 0.2   & \\
         
\hline 

\hline 
\hline 
\end{tabular}
\end{center}
\end{table}

However, there could be different mechanisms capable of delivering larger asteroids. In fact, the numerical simulations performed by \cite{Migliorini1998} show that weaker resonances in the inner belt dominate the delivery process for large asteroids, as they can cause asteroids to increase in orbital eccentricity to Mars-crossing values. These weaker resonances were studied by \cite{NesvornyMorbidelli1998} and \cite{MorbidelliNesvorny1999}. Following \cite{OBrien2005} and \cite{deElia2007}, we treat the dynamical removal of NEAs larger than 5 km from the Inner and Middle belts using a dynamical lifetime of 3.75 Myr, which is the lifetime associated with the Mars-crosser population \citep{Bottke2002}.

We then calculate the number of NEAs in time, coming from a region of the MB, as \citep{OBrien2005}:
\begin{equation}
N_{\text{NEA}}^{r}\left(D,t+\Delta t \right)=N_{\text{NEA}}^{r}\left(D,t\right)+\left| \Delta N_{\text{YE}}^{r}\left(D,t\right)\right| - N_{\text{NEA}}^{r}\left(D,t\right)\frac{\Delta t}{\tau_{\text{NEA}}^{r}}
,\end{equation}
where the second term represents the number of bodies delivered from the MB via the combined action of the Yarkovsky effect and resonances at the time $t$, and the third term is the dynamical removal of NEAs. We do not consider the Yarkovsky effect in the NEA population. In fact, the Yarkovsky effect on the orbits of the NEAs is many orders of magnitude lower than the gravitational perturbations made by the close encounters with terrestrial planets \citep{Granvik2018}. We also do not consider collisions between NEAs and other asteroids. 

\section{Results}
In this section, we present the results of the simulations we performed with the \texttt{ACDC} code. First we describe how we arrive at a satisfying set of initial conditions. Then, we show the final SFDs of the six regions of the MB. Finally, we derive the HFD of the NEAs and find the contribution of the six regions of the MB to the NEAs. 

\subsection{Finding the initial conditions}

\begin{figure*}[h]
    \centering
\includegraphics[width=18cm]{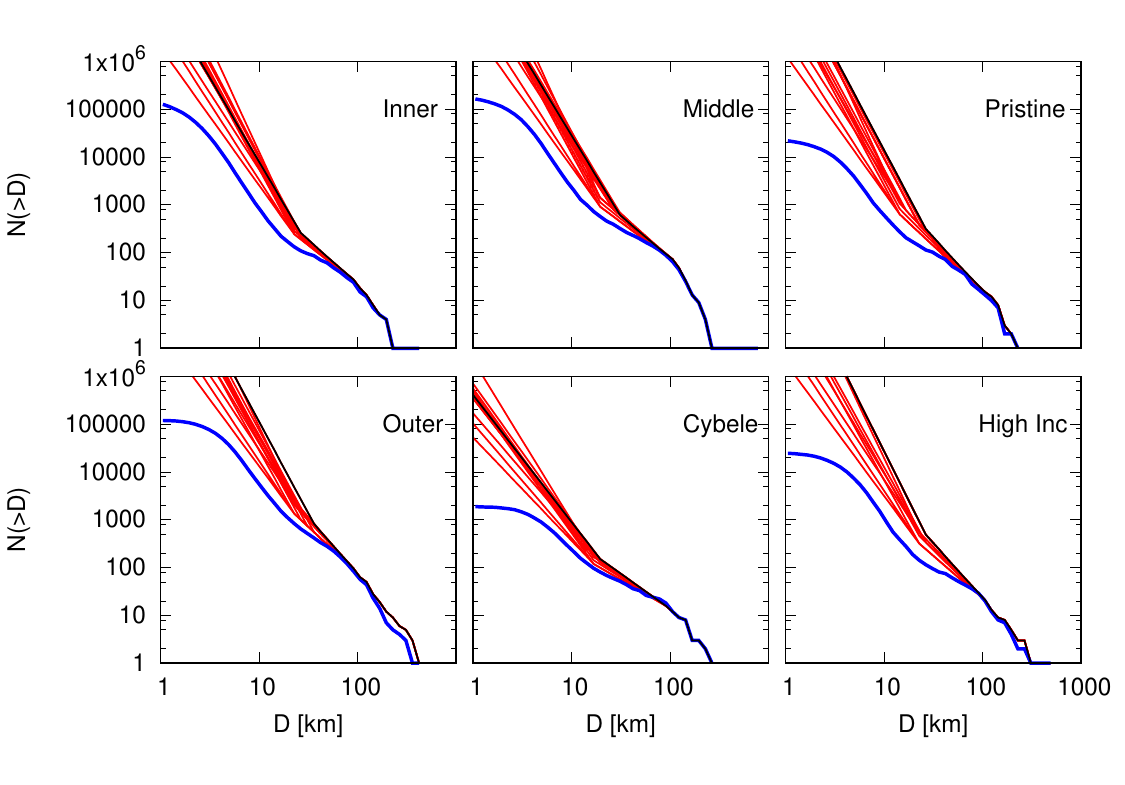}
    \caption{Initial SFDs tested for the six populations of the MB are shown by red lines. Blue lines denote the observed SFDs, while the black lines denote the chosen initial SFDs for this work. A color version of this figure is available in the electronic version of this article.  }
    \label{fig:InicialSFD}
\end{figure*}

The initial conditions in our simulations are the SFDs of the six regions of the MB, which were constructed as three-slope cumulative power laws:
\begin{equation}
N\left( \geq D \right)=\begin{cases}
A_{1}D^{q_{1}} & D \geq D_{1}, \\
A_{2}D^{q_{2}} & D_{2} < D < D_{1}, \\
A_{3}D^{q_{3}} & D \leq D_{2}, 
\end{cases}
\label{eq:Inicial}
\end{equation}
where the free parameters are the cumulative slopes $q_{1}$, $q_{2}$, and $q_{3}$, the cut-off diameters $D_{1}$ and $D_{2}$, and the normalization constants $A_{1}$, $A_{2}$ and $A_{3}$.

We manually added  the parental bodies of observed asteroid families with $D_{\text{PB}}>100$ km and $M_{\text{LR}}/M_{\text{PB}}<0.5$ , which cannot be completely destroyed by collisional evolution \citep{Bottke2005a} or by the Yarkovsky effect \citep{Bottke2001}. We used the parent body diameters estimated by SPH simulations by \cite{Durda2007}, which were summarized by \cite{Broz2013}, and are listed in Table \ref{Tab:Fam}. We calculated mean values in the cases where the estimation takes a wide range of values. In the case of Polana, an asteroid family located in the Inner belt for which there is no available estimation of its parent body size due to its complex structure \citep{Walsh2013,Dykhuis2015}, we considered a parent body size of 115 km, which is consistent with the parent body sizes of other families in the Inner belt. We also manually added the three biggest asteroids: Vesta in the Inner belt, Ceres in the Middle belt, and Pallas in the High-Inclination belt.

\cite{Cibulkova2014} performed a detailed exploration of this free-parameter space in a wide range of slopes and cut-off diameters, and found a set of values that give their best fit with observational data. However, it is very important to note that the main simulations in their work did not include the Yarkovsky effect. By considering the radiation force, the main consequence is a depletion of bodies in the subkilometer range. We found  it necessary to use steeper initial slopes and higher cut-off diameters in order to account for the dynamical removal of bodies and reproduce the observed distributions. 

For the construction of our initial conditions for the SFDs, we explored a wide range of slopes and cut-off diameters by trial and error. The method we used for this exploration is to perform a set of runs for each set of parameter values in order to search for the ones that give lower simultaneous metrics and median SFDs as close as possible to the observed data. The set of initial populations tested, including the one chosen for this work, is plotted in Fig. \ref{fig:InicialSFD}. The slopes and cut-off diameters of the initial conditions used in this work are listed in Table \ref{tab:Inicial}.

\begin{table}
\begin{center}
\caption{Asteroid families in the individual parts of the MB according to \cite{Broz2013}. $D_{\text{PB}}$ is the parent body diameter inferred from SPH simulations \citep{Durda2007}.  We only consider families with $D_{\text{PB}}>100$ km and $M_{\text{LR}}/M_{ \text{PB}}<0.5$. In the case of parent bodies with a wide range of diameters we took a mean value, while in the case of Polana we adopted a parent body size of 115 km. } 
\label{Tab:Fam}
\begin{tabular}{|c c c|}
\hline
\hline
Region & Family & $D_{\text{PB}}$ $\left( \text{km} \right)$  \\
\hline
\hline
Inner    & Erigone & 114 \\
& Eulalia  & 100-160  \\
& Polana  & ? \\
\hline
Middle   & Maria  & 120-130 \\
 & Padua  &  106 \\
 & Misa & 117 \\
 & Dora  &  165 \\
& Merxia  &  121 \\
& Teutonia &  120 \\
& Gefion &  100-150 \\
& Hoffmeister &  134 \\
\hline
Pristine  & Koronis  & 170-180  \\
 & Fringilla  & 130-140 \\ 
\hline
Outer & Themis  & 380-430\\
& Meliboea  &  240-290 \\
& Eos  & 381\\
& Ursula  & 240-280 \\
& Veritas & 100-177\\
& Lixiaohua  & 220 \\
\hline
High-Inclination  & Alauda  & 290-330\\
\hline
\end{tabular}
\end{center}

\end{table}

\begin{table}[h]
\begin{center}
\caption{Input parameters chosen for the initial cumulative size frequency distributions of the six parts of the MB. Here, $q_{1}$ is the slope for bodies bigger than $D_{1}$, $q_{2}$ for bodies with diameters between $D_{1}$ and $D_{2}$, $q_{3}$ for bodies smaller than $D_{2}$, and $A_{1}$ is the normalization constant at $D_{1}$.} 
\label{tab:Inicial}
\begin{tabular}{|c c c c c c c|}
\hline
\hline
Region & $q_{1}$ & $q_{2}$ & $q_{3}$ & $D_{1}$ $\left( \text{km}\right)$ & $D_{2}$$\left(\text{km}\right)$ & $A_{1}$\\
\hline
\hline
Inner    & 3.20  & 1.91 & 3.52 & 90.07   & 23.03 & 20.03\\
Middle & 3.60 & 1.82 & 3.40 & 105.07  & 27.03 & 75.07\\
Pristine & 2.90 & 2.30 & 3.90 & 100.07 & 23.03 & 21.03\\
Outer    & 3.00 & 2.39 & 3.84 & 80.07   &  33.03 &  90.07\\
Cybele  & 1.80 & 1.47 & 2.66 & 80.07   & 17.03  & 17.03 \\
High-Inc.& 3.20 & 2.31 & 4.12 & 100.07 & 23.03 & 30.03 \\
\hline
\end{tabular}
\end{center}
\end{table}

\subsection{Main Belt}

\begin{figure}[h]
    \centering
\includegraphics[width=8cm]{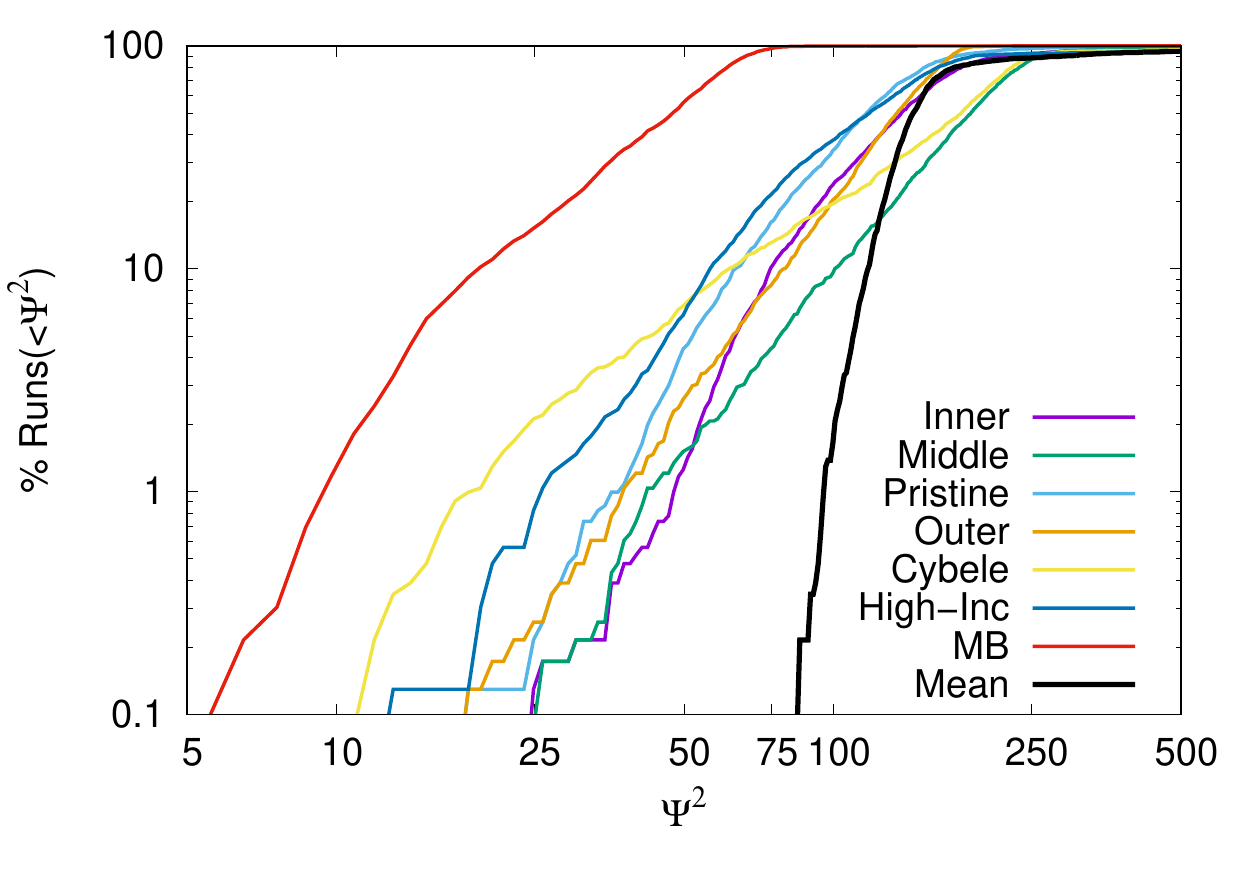}
    \caption{Cumulative percentage distribution of $\psi^{2}$ metrics. We plot $\psi^{2}$ for the six regions of the MB, including a metric for the global MB $\psi^{2}_{\text{MB}}$, and the average  $\psi^{2}_{\text{MEAN}}$  metric. A color version of this figure is available in the electronic version of this article.  }
    \label{fig:Metric}
\end{figure}

\begin{figure}[h]
    \centering
\includegraphics[width=8cm]{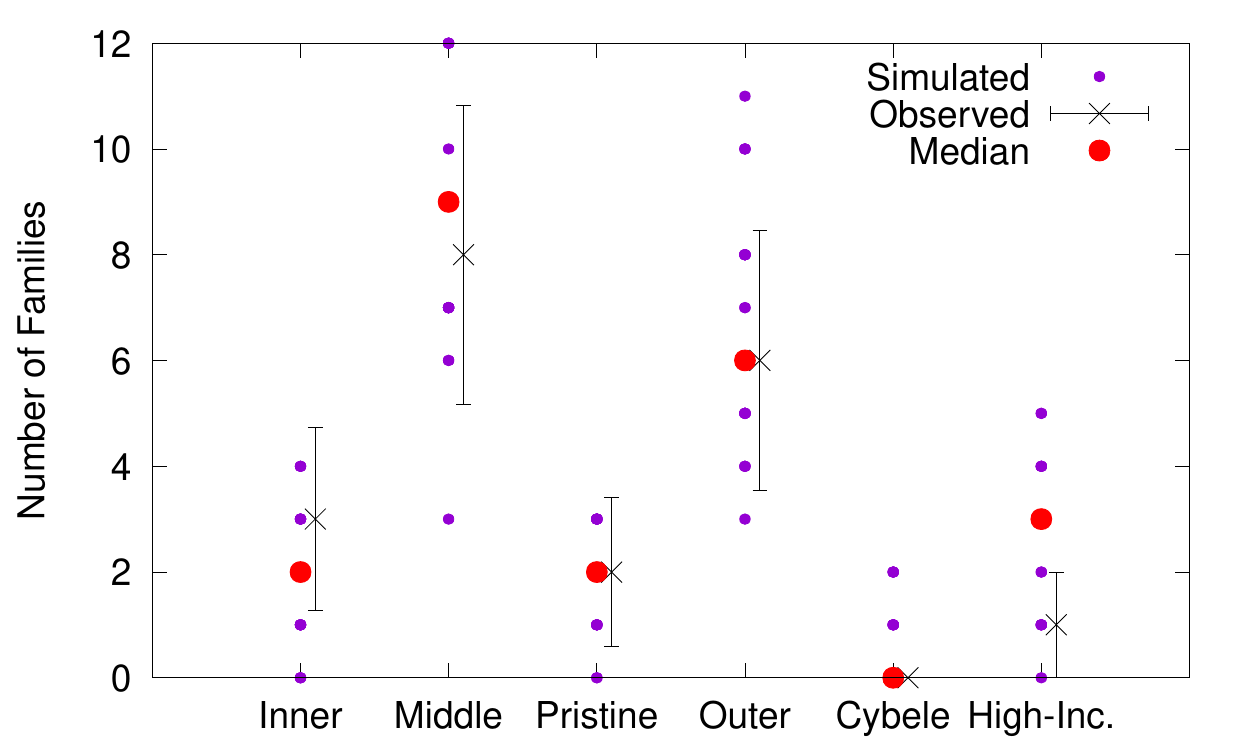}
    \caption{Number of families formed in the individual regions, out of the catastrophic disruption of parent bodies bigger than 100 km and $M_{\text{LR}}/M_{\text{PB}}<0.5$. The error bars denote the uncertainties of the observed numbers of families, calculated as the square root of the observed number of families. }
    \label{fig:Fam}
\end{figure}

\begin{figure*}[ht]
    \centering
\includegraphics[width=18cm]{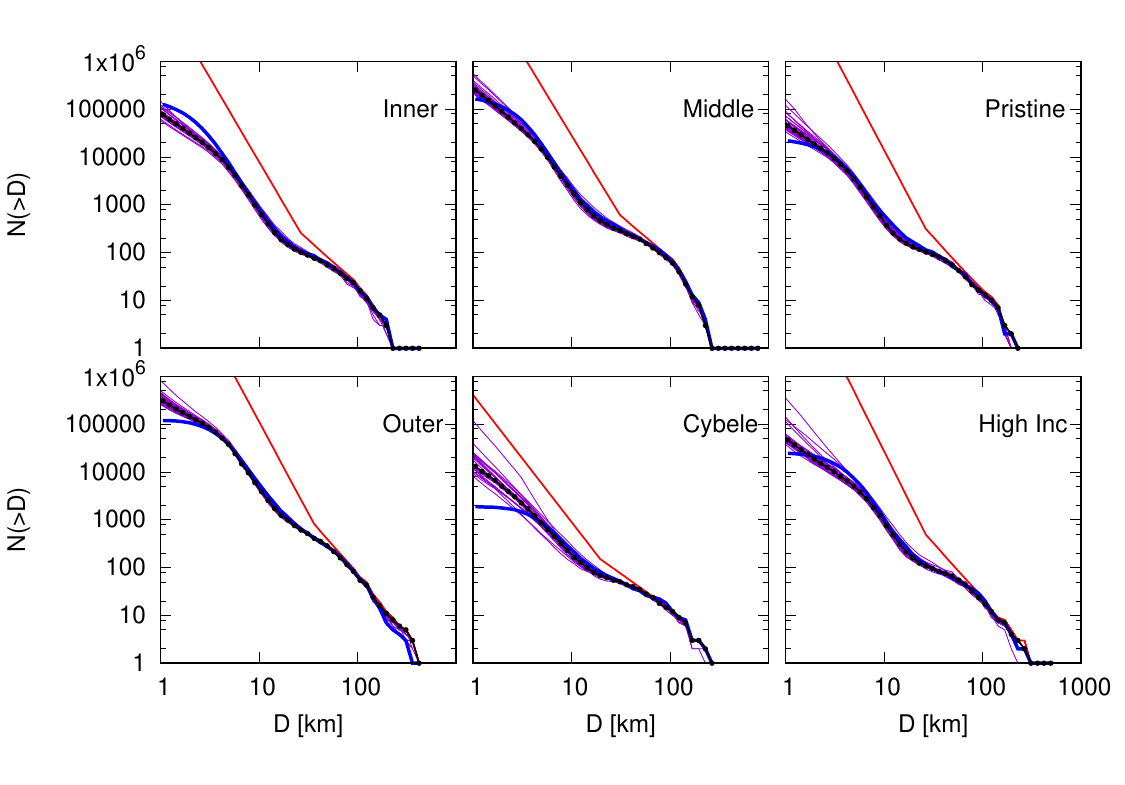}
    \caption{Size--frequency distributions of the six regions of the MB. The selected runs and their median are plotted in purple and black lines, respectively. The observed SFDs are plotted in blue and the initial SFDs are plotted in red. A color version of this figure is available in the online version of this article.}
    \label{fig:FinalSFD}
\end{figure*}

Here, we describe the results of the 2500 runs that we performed using our six-region collisional evolution model regarding the six regions of the MB. The cumulative distributions of the  $\psi^{2}$ metrics of the six regions, along with the global MB  $\psi_{\text{MB}}^{2}$  and the averaged metric $\psi_{\text{MEAN}}^{2} $ of the performed runs are plotted in Fig. \ref{fig:Metric}. We find that all of our runs produce good fits with the global MB. In fact, all of our runs produce $\psi_{\text{MB}}^{2} $ between 5 and 75, and 10\% of runs produce global metrics smaller than 20. The individual metrics give a wider range of values. The lowest metrics of the individual regions lie between 10 and 25.  The regions that produce the lowest metrics are the Cybele and High-Inclination belts, while the Inner and Middle belts give the highest metrics. However, it is worth remembering that, in general terms, not all regions give low metrics simultaneously in a single run. The mean metrics $\psi_{\text{MEAN}}^{2}$, which were calculated by averaging the individual metrics in each run, show values between 80 and 100 and smaller than 120 in 1\% and 10\% of cases, respectively. For our analysis, we select the runs that give mean metrics $\psi_{\text{MEAN}}^{2}$ smaller than 100, which is 1\% of the total. 

\begin{figure}[h]
    \centering
\includegraphics[width=8cm]{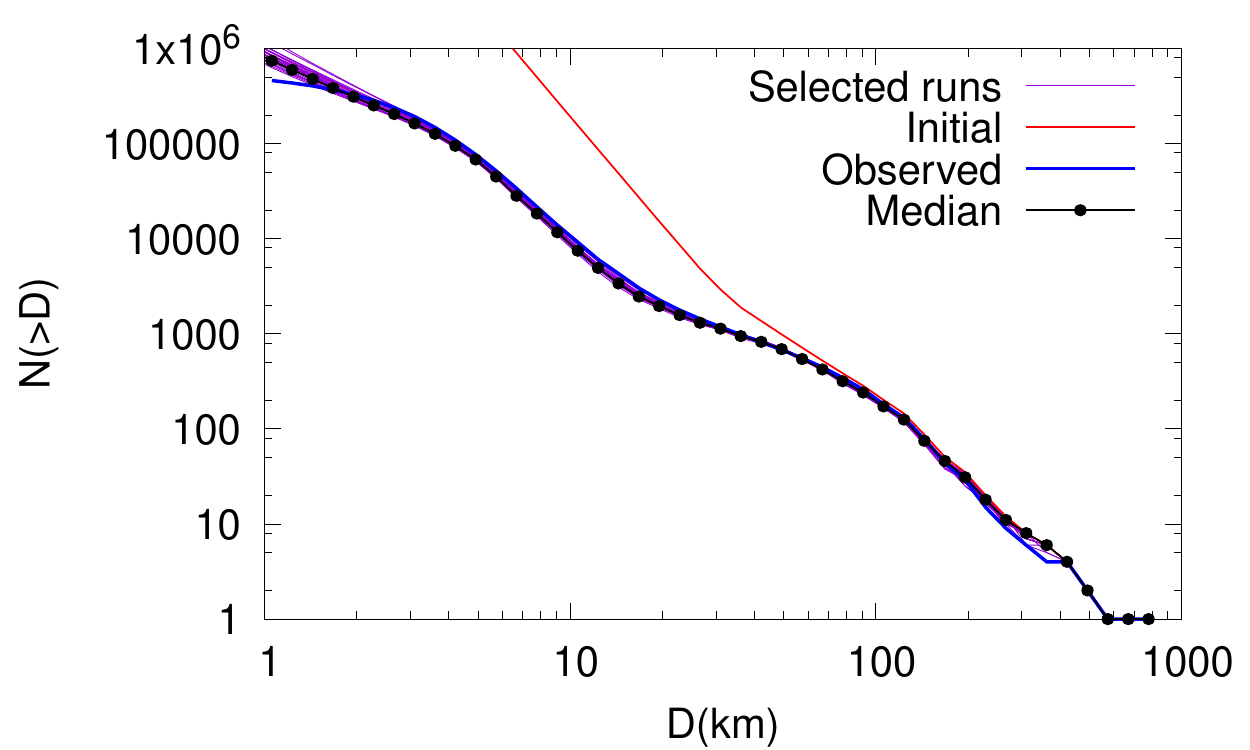}
    \caption{Size--frequency distributions of the global MB, calculated as the sum of the SFDs of the individual regions. The selected runs and their median are plotted in purple and black lines, respectively. The observed SFD is plotted in blue and the initial SFD is illustrated in red. A color version of this figure is available in the online version of this article.  }
    \label{fig:TopMB}
\end{figure}
Our results concerning the formation of asteroid families in the MB is plotted in Fig. \ref{fig:Fam}. We reiterate the fact that we only consider asteroid families produced by the catastrophic disruption of parental bodies bigger than 100 km, such that $M_{\text{LR}}/M_{\text{PB}}<0.5$. Our runs produce a wide number of families, but the median is a good match with the actual observed families. Specifically, the median values lie within the uncertainties, which are calculated as the square root of the observed number of families. The only exception is the High-Inclination region, which has one observed family while our runs produce three families in median. 

The resulting SFDs of the selected runs of the six regions of the MB, along with the median SFDs are plotted in Fig. \ref{fig:FinalSFD}, while the SFD for the global MB is plotted in Fig. \ref{fig:TopMB}. In general terms, we see that our selected runs provide very good fits with the observed data for the six regions and the global MB. Specifically, the results for the Inner, Middle, Pristine, and Outer regions lie in a narrow dispersion along the observed data, while the Cybele and High-Inclination belts show the greatest dispersion between the different runs. We find that the simulations that produce no families in the Cybele region are the ones that produce better fits with observed data.

We find two discrepancies worthy of mention. One is regarding the tail of small bodies of the Inner belts, which is shallower than the observed data. This was also noted in the work of \cite{Cibulkova2014}. To explain this small body deficit of the Inner belt, these latter authors proposed the recent disruption (during the last 100 Myr) of a large parent body that made the actual SFD temporarily steep. An alternative explanation is that the Yarkovsky model in the Inner belts could be making small asteroids drift too quickly. This may be solved in future works by including YORP cycles that result in a slower net Yarkovsky drift, as was shown by \cite{Bottke2015} and \cite{Granvik2017}. The other discrepancy is in the bigger objects of the Outer belt, and is due to the formation of asteroid families. We reiterate the fact that in the construction of the initial conditions, we manually added the estimated parental bodies of the observed asteroid families \citep{Broz2013}. In the case of the Outer belt, these parental bodies have diameters between $\sim$ 100 km and $\sim$ 400 km. We find that our runs reproduce the number of observed families, as shown in Fig. \ref{fig:Fam}. However, the parental bodies that are catastrophically disrupted are mostly smaller than the ones we considered beforehand. This discrepancy may be a consequence of using the same scaling law throughout the entire MB. It was shown in \cite{Cibulkova2014} that the formation of asteroid families depends on the scaling law, and that there could be different scaling laws for different parts in the MB due to the diverse compositions of asteroids \citep{DeMeo2014}. The use of weaker scaling laws in the Outer belt, such as for example the laws for ice in \cite{BenzAsphaug}, may enable a wider range of projectiles capable of catastrophically disrupting these large bodies, increasing the probability of the occurrence of such events, and thus improving the results at the large end. The treatment of these discrepancies represents an interesting starting point for future research.

\subsection{Near-Earth asteroids}
Here, we proceed to describe the results of our simulations for the NEAs. We proceed, as described in Sect. \ref{Sect:neas}, to calculate the evolution of the NEA populations by considering the asteroids delivered via Yarkovsky effect and resonances, and the dynamical remotion due to the many sinks present in the inner Solar System. By doing so, we obtain the current size distribution of NEAs coming from the different source regions. 

Because our simulations originally deal with size bins, we convert the diameters into absolute magnitudes $H$ using the relation \citep{Bowell1989}:
\begin{equation}
D=\frac{1329}{\sqrt{p_{\textrm{V}}}}10^{-H/5}.
\end{equation}
This relationship depends on $p_{\textrm{V}}$, the visual geometric albedo, which is likely size dependent. Indeed, it was suggested by \cite{Mainzer2014a} that there is an increase in albedo with decreasing diameter. In this work, we consider the values $p_{\textrm{V}}=0.11$ for $D>3$ km and $p_{\textrm{V}}=0.15$ for $D<3$ km, which are consistent with the measures of \cite{Mainzer2014a}. 

\begin{figure}[ht]
    \centering
\includegraphics[width=8cm]{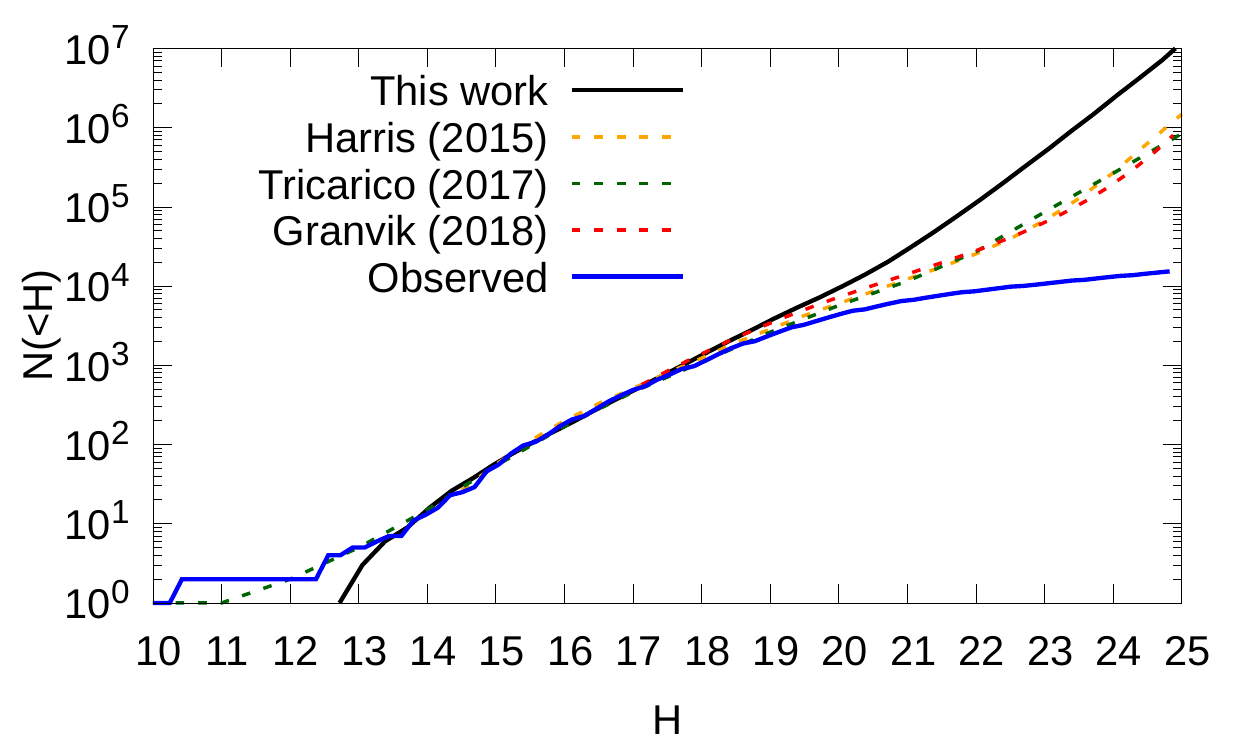}
    \caption{HFD of the NEA population. The median distribution calculated in this work is plotted in black. The observed NEAs, taken from the Minor Planet Center, are plotted in blue.  Dashed lines indicate the derived HFDs of \cite{Harris2015,Tricarico2017,Granvik2018}. A color version of this figure is available in the electronic version of this article.}
    \label{fig:NEAall}
\end{figure}

Recently, \cite{Granvik2018} developed a model of the NEO population that derived a debiased steady-state distribution of orbital elements and absolute magnitudes $H$ in the range $17 < H < 25$  ($ 1.4$ km $< D < 35$ m, for  $p_V = 0.15$). The cumulative magnitude-frequency distribution derived by \cite{Granvik2018} in the range of absolute magnitudes $17<H<25$ is shown in Fig.~\ref{fig:NEAall}. It should be noted that the observed population is thought to be complete up to $H<16$  ($\sim 2$ km for  $p_V = 0.15$) \citep{Tricarico2017}. Figure~\ref{fig:NEAall} shows the total derived NEA population along with the observed NEA population obtained from the Minor Planet Center\footnote{https://minorplanetcenter.net/iau/MPCORB/NEA.txt}. We also plot the recent HFDs derived by \cite{Harris2015,Tricarico2017} for comparison. We find that our median magnitude distribution provides a good match with observations in the range of completeness, $H<16 $. The estimated magnitude distribution agrees with recent works in the range $16<H<20$ ($ \sim 2$ km $< D < 350$ m, for  $p_V = 0.15$), but deviates significantly in $H>20$. The main difference is that our median distribution has a change to a steeper slope around $H\sim 21$  ($\sim 220$ m for  $p_V = 0.15$), while \cite{Granvik2018} and \cite{Harris2015} find a dip in the population in the range $19<H<25$  ($ 550$ m $< D < 35$ m, for  $p_V = 0.15$) . The immediate consequence of this change to a steeper slope in our median magnitude distribution is a higher number of smaller NEAs. Moreover, our work predicts $\sim10^{7}$ asteroids with $H<25$ (35 m, for  $p_V = 0.15$), which is an order of magnitude higher than the $\sim10^{6}$ asteroids of that size made by previous papers. This is most likely caused by the many uncertainties and free parameters regarding the modeling of the Yarkovsky effect, and the exclusion of YORP cycles, especially for smaller asteroids. More specifically, bodies of  $\sim$ 10 m in size have the lowest timescales associated with the Yarkovsky effect, as shown in Fig. \ref{fig:TYE}. As a consequence, this results in a larger number of asteroids delivered to the NEA region. However, since the MB is not complete in the subkilometer range, smaller asteroids are far outside of the observational constraints that we are able to use in this collisional model. Therefore, we limit our analysis to the kilometer range, and will investigate how we can extrapolate results for asteroids bigger than 100 m.

\begin{figure*}[t]
    \centering
\includegraphics[width=8cm]{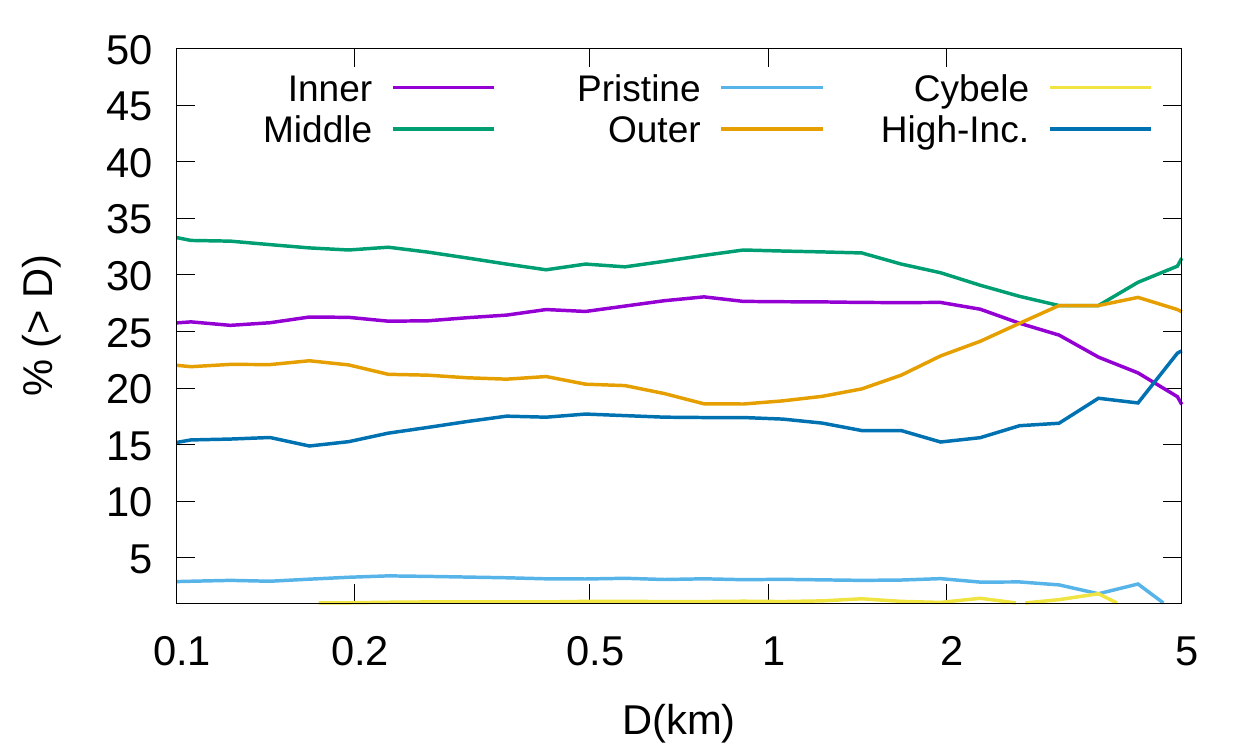}
\includegraphics[width=8cm]{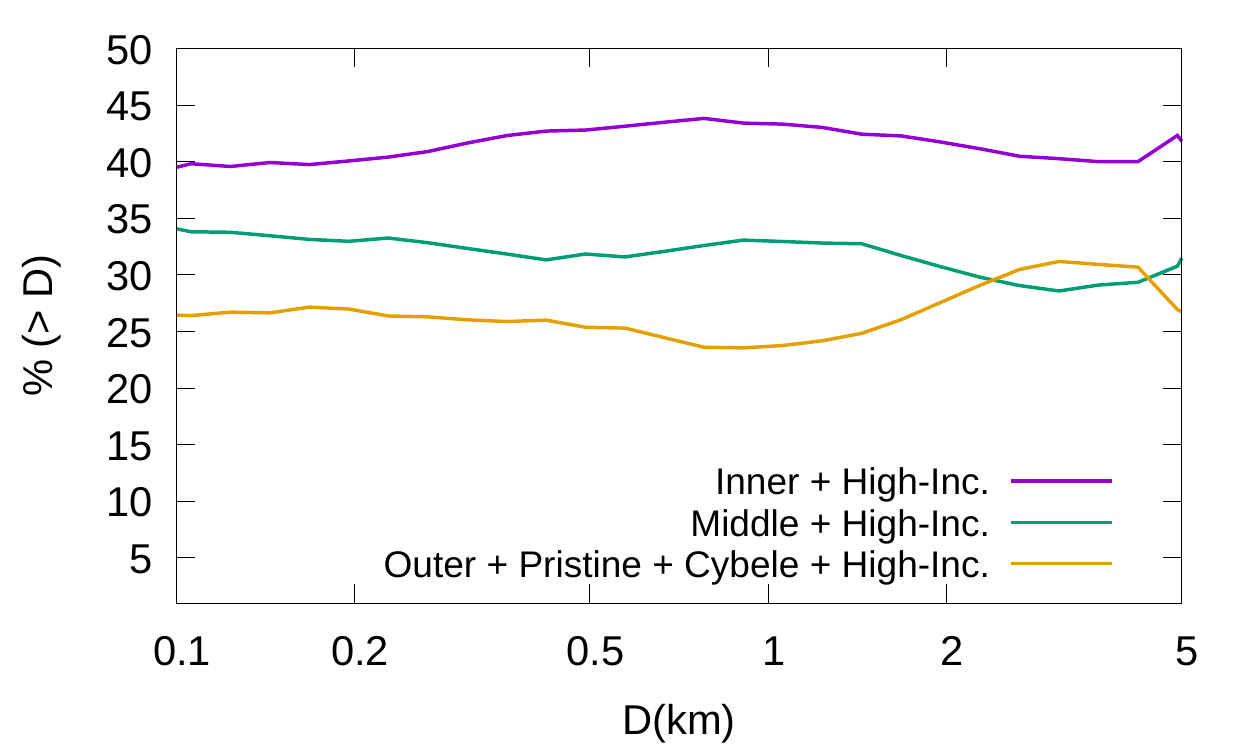}

    \caption{Percentage of NEAs coming from the different regions of the MB with respect to the total cumulative number of NEAs we obtained in the simulations. Left panel: Considering the six regions. Right panel: Distributing the High-Inclination NEAs among the Inner- Middle-, and Outer-belt NEAs, and adding the Pristine and Cybele contributions to the Outer-belt NEAs. A color version of this figure is available in the electronic version of this article.}
    \label{fig:NEApob}
\end{figure*}

We now proceed to studying the contribution of the different regions of the MB to the NEAs. The left panel of Fig. \ref{fig:NEApob} shows the percentages of NEAs coming from the different regions of the MB, which were calculated with respect to the total cumulative number of NEAs we obtained in our simulations. We find that all regions contribute to the NEA population to a greater or lesser extent. The regions that make the smallest contributions are the Pristine and Cybele belts. In fact, these regions beyond the 5:2J resonance provide less than 4\% and 1\% of the NEAs, respectively, in sizes smaller than 5 km. The regions that make the biggest contributions are the Inner and Middle belts, which together account for $\sim60\%$ of small NEAs, providing $\sim27\%$ and $\sim32\%$ of bodies bigger than 1 km, respectively, while the Outer belt provides $\sim18\%$. We see that the percentages remain similar in smaller sizes. We obtain that the High-Inclination belt is the source of $\sim17\%$ of NEAs larger than 1 km. In our partition of the MB, the High-Inclination region is made of all the asteroids with inclinations greater than $20\degr$, spanning the entire MB in semi-major axis. Thus, the High-Inclination region contains asteroids that could be assumed to be part of the Inner, Middle, and Outer belts if we consider the whole range of inclinations. If we redistribute the contribution of the High-Inclination region into the Inner, Middle, and Outer belts with respect to the way these are defined in section \ref{Sect:neas}, and adding the Pristine and Cybele NEAs as part of the Outer region, we see in the right panel of Fig. \ref{fig:NEApob} that the results change, as follows: we find that $\sim 77\%$ of the NEAs come from the Inner and Middle belts, which provide $\sim44\%$ and $\sim33 \%$, respectively, and $\sim23\%$ come from the Outer belt. These values are valid for NEAs bigger than 1 km spanning the whole range of inclinations, but we see in Fig. \ref{fig:NEApob} that the percentages remain similar in smaller sizes. In the multikilometer range, the Inner belt is the main source of NEAs, as the right panel in Fig. \ref{fig:NEApob} shows. We note that the share of Outer belt NEAs slightly increases, matching the Middle belt contribution in $\sim 3$ km.  However in larger size ranges, as is explained below, the number of NEAs becomes too small for this to be taken as a strong statistical result.

\begin{figure}[htp]
    \centering
\includegraphics[width=8cm]{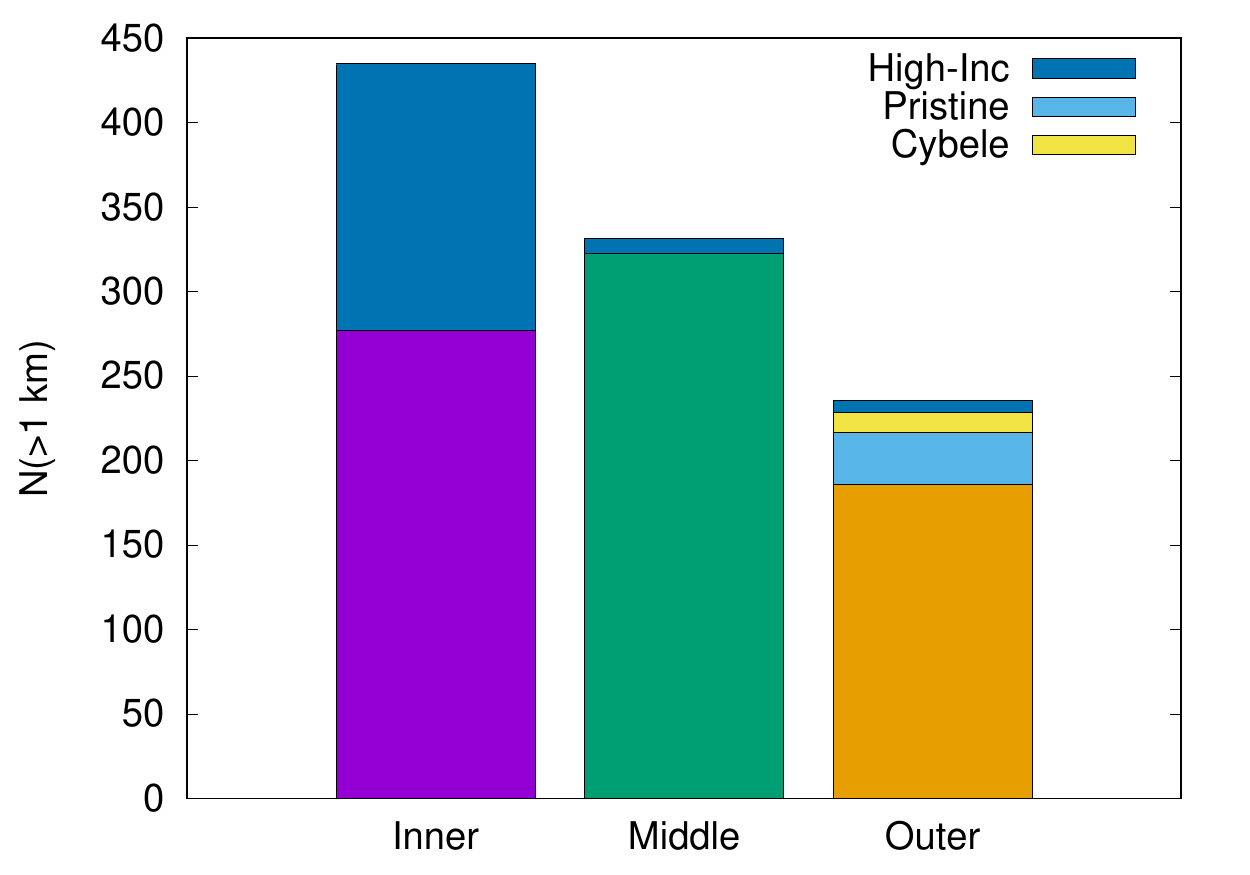}
\includegraphics[width=8cm]{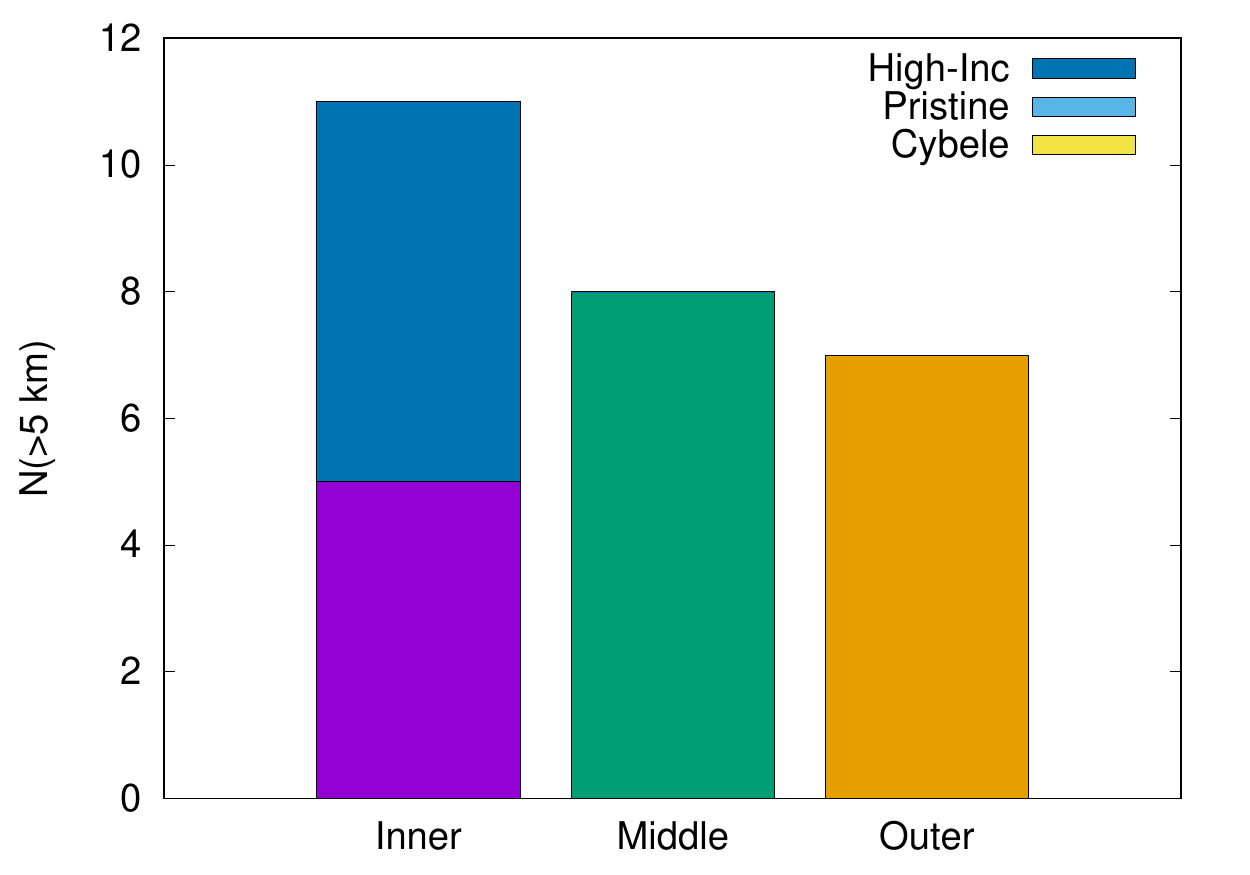}

    \caption{Number of NEAs bigger than 1 km (top) and 5 km (bottom), according to their source regions. The small contributions of the Pristine and Cybele belts are added to the Outer belt, while the High-Inclination NEAs are distributed among the Inner, Middle, and Outer NEAs. A color version of this figure is available in the electronic version of this article. }
    \label{fig:Histogramas}
\end{figure}

The top and bottom panels of Fig. \ref{fig:Histogramas} show the number of NEAs bigger than 1 km and 5 km, respectively, according to their source regions; the total numbers we obtain in our simulations are 1002 and 26, respectively. We find that the Inner region provides 432 NEAs bigger than 1 km, out of which 277 come from the proper Inner belt, while 158 come from the corresponding section of the High-Inclination region. Similarly, 323 NEAs come from the Middle belt with 9 from the Middle High-Inclination region. The Outer region and Outer High-Inclination belt provide  186 and 8 NEAs bigger than 1 km, respectively, while the Pristine and Cybele belts provide 31 and 12, respectively. The bottom panel of Fig. \ref{fig:Histogramas} shows that the Middle and Outer belts provide 8 and 7 NEAs bigger than 5 km, respectively. We see that the High-Inclination region does not provide large NEAs corresponding to the Middle and Outer belts, and neither the Pristine nor Cybele belt provides any NEAs in this size range. Moreover, we find that, out of the 11 NEAs bigger than 5 km that come from the Inner part, 6 came from the High-Inclination belt. 

It is interesting to compare the results concerning NEAs coming from the High-Inclination belt with the actual observed distribution of NEAs with high inclinations (i.e., $i> 20\degr$). As we show in Fig.~\ref{fig:NEApob}, the percentage of NEAs bigger than 1 km coming from the High-Inclination belt is $\sim$ 17\%. However, by exploring the MPC database, we find that $\sim$ 108 NEAs with $H <$ 16 (the complete population) have orbital inclinations $i \gtrsim$ 20$^{\circ}$, which is $\sim$ 56\% of the total number of NEAs with $H <$ 16. The existence of high-inclination NEAs coming from a source other than the High-Inclination belt is something this study is not able to explain as we do not calculate the evolution of the orbital elements. However, the simulations performed by \cite{Granvik2016,Granvik2018} show that there are dynamical processes that modify the inclinations of asteroids when leaving the MB. In particular, \cite{Granvik2016} show that 3:1J, 5:2J, and 7:3J resonances, and especially the 2:1J resonance, are able to deliver low-inclination asteroids to the NEA region with inclinations beyond 20\degr. Moreover, these mean-motion resonances can also decrease inclinations in such a way that asteroids that initially had high inclinations enter the NEA region with inclinations smaller than 20\degr. Furthermore, \cite{Granvik2018} show that NEAs delivered from these latter four resonances and also from $\nu_{6}$ and 3:1J are able to increase their inclinations inside the NEA region through dynamical evolution. Therefore, the High-Inclination belt is not necessarily the source of NEAs with large inclinations. Moreover, the High-Inclination belt is the source of only a fraction of the observed NEAs with high inclinations, while the rest increased their inclinations through dynamical evolution.

It is interesting to compare our results regarding the proportions of NEAs from different source regions with previous works. However, comparisons between collisional and dynamical simulations of the MB must be handled with care. Indeed, it is not yet possible to address the collisional and fragmentation processes simultaneously with the dynamical evolution in the MB  in a self-consistent way, and therefore some differences are expected. 

 \cite{Bottke2002} found that $\sim 60\%$ of NEOS come from the Inner belt, $\sim 25\%$ from the Middle belt, $\sim8\%$ from the Outer belt, and $\sim6\%$ for the Jupiter-family Comets, which we do not consider here. The contribution from the Outer belt in our work, adding the NEAs provided by the Pristine, Cybele, and the corresponding fraction of High-Inclination, is $\sim 25\%$ . However, the lifetime in the NEA region of bodies from the Outer belt found by \cite{Bottke2002} is 0.14 Myr, much smaller than the value derived in our work by averaging the lifetimes of 0.68 Myr and 0.4 Myr, associated with the 5:2J and 2:1J resonances, respectively \citep{Granvik2018}. If we use the time of 0.14 Myr for the Outer belt, we find that the percentages of the total NEAs found in the Inner, Middle, and Outer belts are $\sim$50\%,$\sim$ 40\%, and $\sim$10\% respectively, which are closer to the percentages found by \cite{Bottke2002}. However, using this latter lifetime for the Outer-belt NEAs results in a HFD that does not match the observed data. 

It was shown by \cite{Granvik2018} that $\nu_{6}$ is the resonance that provides the highest number of NEAs throughout the whole size range, followed by 3:1J, which immediately results in a dominance of the Inner belt as the main source of NEAs. As shown above, we are able to obtain a similar result when we redistribute the contribution of the High-Inclination region between the Inner, Middle, and Outer NEAs.

\section{Conclusions and discussion}

Here, we present the \texttt{ACDC} code, a six-part collisional and dynamical evolution model of the MB, which we split into six regions according to the positions of the 3:1, 5:2, 7:3, and 2:1 mean-motion resonances with Jupiter and the $\nu_{6}$ secular resonance with Saturn. The six regions we consider, following the previous work of \cite{Cibulkova2014}, are the Inner, Middle, Pristine, Outer, Cybele, and High-Inclination belts. The \texttt{ACDC} code calculates the evolution in time of the number of objects in each part of the MB due to collisions between the asteroids of the different regions. \texttt{ACDC}  is a statistical code, as it treats big impact events with a random number generator and Poisson statistics. It also considers the dynamical depletion of the MB via the Yarkovsky effect, which slowly drifts small asteroids into the resonances and delivers them into the NEA region. We used the \texttt{ACDC} code to derive SFDs of the six regions of the MB, estimate a HFD for the NEAs, and studied the contribution of each region to the NEA population. 

The results of this work can be summarized as follows:
\begin{itemize}

\item We find that our six-part collisional evolution model shows a good fit to the observed data in the six regions of the MB, and to the MB considered as a single population.  We are also able to reproduce, within the expected uncertainties, the number of observed asteroid families formed from the catastrophic disruption of parent bodies bigger than 100 km. There are two main discrepancies in the resulting SFDs. The first is located at the small end of the SFD for the Inner belt, which may be corrected in future works with a better model of the Yarkovsky effect.The second is related with the largest bodies in the Outer belt, and may be corrected by exploring different scaling laws in that particular region which may enable a wider range of projectiles capable of disrupting these large bodies. The treatment of these discrepancies represents an interesting starting point for future research. 
\item  We obtain a good fit with the currently observed magnitude distribution of NEAs in the range of completeness, $H<16$, and with the recent estimations in the range $17<H<20$. Our HFD deviates from previous estimates for smaller bodies. This discrepancy is most likely caused by the many uncertainties and free parameters regarding the current model of the Yarkovsky effect, especially for smaller asteroids. Further studies may require an improved model of the Yarkovsky effect for smaller asteroids, and may also consider the inclusion of the YORP effect, which was not considered here. However, the subkilometer-sized asteroids are out of the available observational constraints for the present collisional model, as the MB is not complete below the kilometer scale.
\item We find that all the regions in the MB contribute to the NEA population to a greater or lesser extent. The most important sources are the Inner and Middle belts followed by the Outer belt. The Pristine and Cybele belts make a very small contribution to the NEAs.
\item It is not possible from this work to address the origin of NEAs with inclinations greater than 20\degr, as the resonances in the MB and the dynamics in the NEA region are able to increase and decrease the inclinations of the asteroids \citep{Granvik2017,Granvik2018}. The High-Inclination belt can be the source of only a fraction of the currently observed NEAs with large inclinations, while the rest of the NEAs obtained their inclinations through dynamical evolution. Further dynamical studies may be able to provide more clarity on the origin of high-inclination NEAs.  
\end{itemize}

We have improved on previous collisional studies as we have consistently included a general model of the Yarkovsky effect into a six-population collisional evolution model of the MB for the first time. Moreover, our work forms a starting point for further studies and applications regarding the collisional history of the MB, and also any other small-body population in the Solar System. We also obtained the population of NEAs, identifying their proportion from each source in the MB. This is  undoubtedly an important contribution concerning the composition and diversity we can expect in such objects.

\begin{acknowledgements}

The present investigation work was partially financed by Agencia Nacional de Promoci\'on Cient\'ifica y Tecnol\'ogica (ANPCyT) through PICT 201-0505, and by Universidad Nacional de La Plata (UNLP) through PID G144. We acknowledge the financial support by Facultad de Ciencias Astron\'omicas y Geof\'isicas de La Plata (FCAGLP) and Instituto de Astrof\'isica de La Plata (IALP) for extensive use of their computing resources. We are grateful to Miroslav Brož for kindly providing us the observed data, and to Mikael Granvik (as a reviewer) for valuable comments and suggestions that helped us improve the manuscript.
\end{acknowledgements}

\bibliographystyle{aa} 
\bibliography{Zain2020} 

\end{document}